\newcommand{\pa}{\partial}
\newcommand{\beq}{\begin{equation}}
\newcommand{\eeq}{\end{equation}}
\newcommand{\be}{\begin{equation}}
\newcommand{\ee}{\end{equation}}
\newcommand{\bea}{\begin{eqnarray}}
\newcommand{\eea}{\end{eqnarray}}
\newcommand{\ena}{\end{eqnarray}}
\newcommand{\bear}{\begin{eqnarray}}
\newcommand{\ear}{\end{eqnarray}\noindent}

\newcommand {\non}{\nonumber}
\newcommand{\la}{\langle}
\newcommand {\ra}{\rangle}

\newcommand{\cF}{{\cal F}}

\newcommand{\s}{\sigma}
\renewcommand{\t}{\tau}
\renewcommand{\d}{\delta}
\newcommand{\tq}{\widetilde{q}}

\newcommand{\rmd}{{\rm d}}

\newcommand{\bt}[1]{{\bar t}}
\newcommand{\ts}{\textstyle}
\newcommand{\ea}{\!\!\! = \!\!\!}
\newcommand{\half}{{\ts \frac{1}{2}}}
\documentclass[11pt]{article}
\input epsf
\usepackage{amsmath,amsfonts,epsfig,color,latexsym}
\usepackage{amssymb}
\renewcommand{\r}{\rho}
\renewcommand{\s}{\sigma}

\csname @addtoreset\endcsname{equation}{section}
\begin{document}
\begin{titlepage}
{\hbox to\hsize{ \hfill {AEI--2003--039}}} {\hbox to\hsize{${~}$\hfill
{Bicocca-FT-03-13}}}
{\hbox
to\hsize{${~}$ \hfill {IFUM--758--FT}}}
{\hbox to\hsize{${~}$ \hfill {LAPTH-980/03}}}

\begin{center}
\vglue .06in {\huge\bf Four-point correlators of BPS operators 
\\ in ${\cal N}=4$ SYM at order $g^4$}
\\[.3in]
{\large {$\rm{G. Arutyunov}^{(a)}$}}, {\large{$\rm{S. ~Penati}^{(b)}$,
$\rm{A. ~Santambrogio^{(c)}}$, $\rm{E.
~Sokatchev^{(d)}}$}}
\\[.3in]
\small

$^{\rm(a)}$ {\it Max-Planck-Institut f\"ur Gravitationsphysik,
Albert-Einstein-Institut, \\
Am M\"uhlenberg 1, D-14476 Golm, Germany}
\\
[.03in]
$^{\rm(b)}${\it Dipartimento di Fisica, Universit\`a degli studi di
Milano-Bicocca and INFN, Sezione di Milano, piazza della Scienza 3,
I-20126 Milano, Italy}
\\
[.03in] $^{\rm(c)}${\it Dipartimento di Fisica, Universit\`a degli studi di
Milano and INFN, Sezione di Milano, via Celoria 16, I-20133 Milano, Italy}
\\
[.03in] $^{\rm(d)}${\it Laboratoire d'Annecy-le-Vieux de Physique Th\'{e}orique\footnote[1]{UMR 5108 associ\'{e}e \`{a}
 l'Universit\'{e} de Savoie}
LAPTH, B.P. 110, F-74941 Annecy-le-Vieux, France}
\\[.3in]
\normalsize


{\bf ABSTRACT}\\[.0015in]
\end{center}

We study the large $N$ degeneracy in the structure of the four-point
amplitudes of $\half$-BPS operators of arbitrary weight $k$ in perturbative 
${\cal N}=4$ SYM theory. At one loop (order $g^2$) this degeneracy manifests
itself in a smaller number of independent conformal invariant functions
describing the amplitude, compared to AdS$_5$ supergravity results. 
To study this 
phenomenon at the two-loop level (order $g^4$) we consider a particular ${\cal N}=2$ hypermultiplet
projection of the general ${\cal N}=4$ amplitude. Using the formalism of ${\cal N}=2$ harmonic superspace we then explicitly compute this four-point correlator at two loops and identify the corresponding conformal invariant functions.

In the cases of  $\half$-BPS operators of weight $k=3$ and $k=4$ the one-loop large $N$ degeneracy is lifted by the two-loop corrections. However, for weight $k > 4$ the degeneracy is still there at the two-loop level. This behavior suggests
that for a given weight $k$ the degeneracy will be removed if perturbative corrections of sufficiently high order are taken into account.
These results are in accord with the AdS/CFT duality conjecture.

\vskip 70pt
${~~~}$ \newline

\end{titlepage}

\setcounter{footnote}{0}

\section{Introduction}

In recent years the supersymmetric quantum conformal field theories in dimensions higher than two have undergone a renaissance due to the discovery of the AdS/CFT correspondence \cite{M}. In the first place this concerns  ${\cal N}=4$ super Yang-Mills (SYM) theory in four dimensions whose conjectured dual is type IIB (superstring) supergravity on an anti-de Sitter background.

On the string theory side the present investigations are focused either on low-energy supergravity or on semiclassical string solutions carrying non-vanishing global charges of the superconformal group.
In the latter case a direct comparison of the corresponding
string energies to the spectrum of scaling dimensions
of certain composite operators in ${\cal N}=4$ SYM is possible \cite{P,BMN,GKP}.

According to the duality conjecture, AdS supergravity is dual 
to the gauge theory
taken at an infinite value of the 't Hooft coupling $\lambda$ and in the large $N$
limit, where $N$ is the rank of the gauge group (S)U($N$).
In this limit the gauge theory reveals the following peculiar behavior:
\begin{itemize}
\item the sector of protected operators is enlarged;
\item the spectrum of anomalous dimensions degenerates;
\item the symmetry is enhanced.
\end{itemize}

The first property is a simple consequence of the rather general fact that
in the large $N$ limit the anomalous dimension
of a multi-trace operator made of single-trace constituents
is equal to the sum of their individual anomalous dimensions.
In this sense, in the large $N$ limit the protected operators form
a ring\footnote{When computing the two-point function of a multi-trace operator, one can readily see
that the leading $1/N$ contribution is due to the Feynman graphs which contribute
separately
to the two-point functions of the individual single-trace constituents.}.

The second property is more subtle and it is reminiscent of the free theory pattern, where usually several operators with the same canonical dimension coexist.
It is well known that protected (for instance, $\half$-BPS) operators with the same quantum numbers can be realized in the gauge theory both as single- and multi-trace operators.
Thus, multiplying protected operators with the same quantum numbers
by an unprotected operator, we obtain different operators with the
same dimension in the planar limit, hence the degenerate spectrum. Such
a mechanism is essentially due to the conformal supersymmetry responsible
for the protection and, again, due to the large $N$ factorization
property of the Feynman graphs.
However, this is not the whole story. As was recently shown in Ref. \cite{BKS},
there exist, for instance, single-trace operators (which cannot be obtained as the product of a protected and an unprotected operators) whose
anomalous dimensions coincide up to two loops. This is evidence for an extra symmetry in the planar limit which is responsible for this degeneracy of the perturbative spectrum. The symmetry enhancement \cite{BKS} is an interesting property of the gauge theory related, for instance, to the integrability of the spin chain Hamiltonian
\cite{MZ}. The latter can be identified with the one-loop dilatation operator in ${\cal N}=4$ SYM.

Recently, another interesting peculiarity of the large $N$ limit of ${\cal N}=4$ SYM
was found \cite{AS}. We recall that in regard to the supergravity approximation of the gauge theory, 
the most interesting object to study are the four-point correlation functions 
of  $\half$-BPS operators (for reviews see, e.g. \cite{AGMOO,DHF}). These operators are dual to the Kaluza-Klein states of the compactified type IIB supergravity. They form isolated series of superconformal representations and their scaling dimension as well as their two- and three-point correlation functions are protected from receiving quantum corrections. In contrast, their four-point correlators undergo  perturbative renormalization.

The four-point amplitude of $\half$-BPS operators with arbitrary weights (dimensions) $k_p$, $p=1,2,3,4$ is specified by a number of conformal invariant functions depending on the two conformal cross-ratios. There is a general procedure based on the field-theoretic insertion formula which allows one to determine the maximal number of {\it a priori} independent such functions appearing in the quantum part of the amplitude \cite{ADOS,HH}.

The explicit amplitudes corresponding to the cases where all $k_p\equiv k$ are equal to 2, 3 or 4 have been computed in the supergravity approximation by using the effective five-dimensional description of type IIB, and also in perturbation theory at one loop (order $g^2$) and in the planar limit. It was then observed in Ref. \cite{AS} that the perturbative amplitudes exhibit a certain degeneracy in comparison to their supergravity counterparts. Starting from $k=4$, the number of conformal invariant functions describing the amplitude at one loop is smaller than in the supergravity regime. We can formulate this fact as another interesting property of the large $N$ limit:
\begin{itemize}
\item 
the perturbative four-point amplitudes of $\half$-BPS operators  degenerate.
\end{itemize}
Thus, not only the two-point functions (the spectrum) but also the higher-point
perturbative amplitudes exhibit simplified features in the large $N$ limit.

The aim of the present paper is to the study this degeneracy in more detail. Recall that for the first time the degenerate situation occurs  for $k=4$. In the supergravity regime one finds two different conformal invariant functions, precisely the number predicted on general grounds. However, at one loop the two functions describing the large $N$ amplitude coincide. It should be noted that the perturbative computation has been performed for $\half$-BPS operators realized as {\it single-trace} operators. Generically, the $\half$-BPS operators can be realized as mixtures of single- and multi-trace operators
with arbitrary mixing coefficients.
Therefore, two natural questions can be asked about this large $N$ degeneracy:
{\it
\begin{enumerate}
\item Does the operator mixing influence the degeneracy?
\item Is the degeneracy lifted when the higher-loop corrections are
taken into account?
\end{enumerate}
}

We study the first question by using the $k=4$ amplitude as the simplest example.
The $\half$-BPS operator of dimension 4 is a mixture of one single- and one double-trace operator, the latter is made of two single-trace constituents of dimension 2. Firstly, we show that the free  amplitude involving mixed operators coincides with that for single traces in the large $N$ limit, provided that the mixing parameter scales faster than $1/\sqrt{N}$. As discussed in Ref. \cite{AS}, the supergravity-induced amplitude admits {\it a unique splitting} into a ``constant'' and an ``interacting'' parts. The former precisely matches the free amplitude computed for single-trace operators. In the AdS/CFT language the interacting part must comprise all loop corrections to the free amplitude. Clearly, in order not to affect this agreement with the constant part of the supergravity amplitude, in field theory we should allow only mixing which is more heavily suppressed than $1/\sqrt{N}$. Secondly, we argue that with such a mixing the contribution of the double-trace operators at one loop is suppressed in the large $N$ limit as well, so the degeneracy is not affected by the operator mixing.


We then investigate the issue of the higher-loop corrections by extending the
perturbative treatment to two loops (order $g^4$). To illustrate the more general picture, we consider the amplitude involving $\half$-BPS operators of equal but
arbitrary weight $k$. By performing an explicit diagrammatic computation we find that
at two loops the degeneracy occurring in the case $k=4$ is lifted:
Two different conformal invariant functions emerge. However, for amplitudes involving operators of weight $k > 4$ the degeneracy persists even at the two-loop level. These observations indicate that for a fixed weight $k$ the degeneracy is removed when loop corrections of sufficiently high order (depending on $k$) are taken into account.
Since the supergravity-induced amplitudes comprise all perturbative corrections, it is natural to expect them to realize the maximal  possible number of different conformal invariants.

In fact, the perturbative degeneracy phenomenon has a rather simple origin. At a given order of perturbation theory we have only a finite number of topologically different planar interacting graphs. Whether all of them are indeed present depends on the weight $k$ of the four composite operators for which the amplitude is computed. It is clear that increasing $k$ makes all possible interacting topologies appear. Eventually, we reach the point of saturation beyond which increasing the weight amounts to decorating the interacting graphs with free propagators. In the large $N$ limit this does not generate new conformal invariants. On the contrary, fixing $k$ and increasing the order of perturbation theory ``awakes" new interacting topologies and hence generates new conformal invariants, thus lifting the degeneracy.

One could also ask if there is any relationship between the degeneracy of the four-point amplitudes and that of the spectrum observed in Ref. \cite{BKS}. Any conformal invariant function can be expanded over an infinite basis of conformal partial wave amplitudes, each of which represents the contributions of an individual primary operator. The degeneracy of the four-point functions is therefore related to some kind of degeneracy of the spectrum and of the OPE coefficients.
It would be interesting to find out if
the degeneracy of the spectrum is also removed when higher loop corrections are taken into account. Indeed, in the lattice picture \cite{MZ} the $L$-loop dilatation matrix induces the mixing of the $L+1$ neighboring elementary fields constituting a composite operator, i.e. with $L$ growing the effective interaction is spreading over the lattice and can subsequently increase its size which is equal to the dimension of the operator. When this happens, the degeneracy might be lifted in a way similar to that for the four-point amplitudes.

The organization of the paper is as follows. In Section 2 we describe the structure of the  ${\cal N}=4$ four-point amplitude of  $\half$-BPS operators determined by the field-theoretic insertion procedure and discuss its ``pure'' ${\cal N}=2$ hypermultiplet projection. It involves a subset of $k-2$ conformal invariant functions (or one such function for $k=2$) from the original ${\cal N}=4$ amplitude. For $k>4$ this subset is not complete but it is sufficiently large to illustrate the degeneracy problem. The advantage of the hypermultiplet projection is that we can use the quantum formalism of ${\cal N}=2$ off-shell harmonic superspace \cite{Galperin:1984av} which is particularly efficient at two loops \footnote{Correlation functions of $\half$-BPS operators at order $g^4$ have also been computed in the ${\cal N}=1$ superspace formalism \cite{PSZ,BKRS}.}. For the sake of clarity, at the end of Section 2 we present and discuss our main results, whose derivation is explained in detail in Sections 4 and 5. In Section 3 we discuss the influence of operator mixing on the degeneracy problem. In Section 4 we summarize the ${\cal N}=2$ insertion procedure and the Feynman diagram tools necessary to perform the two-loop computation of the hypermultiplet projection. In Section 5 a diagrammatic calculation is presented and the corresponding conformal invariants are identified. Finally, in Section 6 we use the operator product expansion and the knowledge of the one-loop anomalous dimensions of certain operators to get an insights into the structure of the $k=3$ and $k=4$ amplitudes, independently of the diagrammatics. In particular, for $k=3$ we are able to completely reconstruct the two-loop amplitude from these OPE considerations. Appendix A clarifies the technical tool of harmonic analyticity which we systematically use in order to drastically reduce the number of relevant Feynman graphs. Appendix B contains the details of the graph calculation for $k=3$ and 4 whose generalization for arbitrary $k$ is given in Section 5.

\section{Generalities and main result}

The lowest component of a $\half$-BPS multiplet in ${\cal N}=4$ SYM is a real scalar field of dimension $k$ transforming in the irrep $[0,k,0]$ of the R symmetry group $SO(6)\sim SU(4)$.  In terms of the elementary fields it can be realized, e.g., as a single-trace operator
\bea
\label{bps'}
\mbox{Tr}(\phi^{\{a_1}\ldots \phi^{a_n\}})\,.
\eea
Here $\phi^a$, $a=1,\ldots, 6$ are the ${\cal N}=4$ SYM scalars and
$\{,\}$ denotes  traceless symmetrization.
A convenient way to handle the SO(6) indices is to project the operator (\ref{bps'}) onto the highest weight state of the irrep $[0,k,0]$. This can be done with the help of a complex null vector $u^a$ ($u^au^a=0$) carrying U(1) charge, which can be viewed as a harmonic variable parametrizing the coset space SO(6)/SO(2) $\times$ SO(4) (see, e.g., Ref. \cite{ADOS} for details):
\bea
\label{bps}
{\cal O}^{(k)}= u^{a_1}\ldots u^{a_k}\mbox{Tr}(\phi^{a_1}\ldots \phi^{a_n})\, .
\eea

We start by summarizing the general properties of the four-point amplitude of
$\half$-BPS operators in the ${\cal N}=4$ SYM theory. It is
sufficient to restrict our discussion to the case where all the operators
involved have equal weights $k$. On the basis of conformal and R symmetry alone the general form of the four-point amplitude
can be parametrized as follows:
\bea
\label{gampl}
\la k|k|k|k \ra=\sum_{m+n+l=k}a_{mnl}^{(k)}(s,t)
{\cal X}^m{\cal Y}^n{\cal Z}^l \, ,
\eea
where $m,n,l$ are non-negative integers. This expression is a polynomial in the three elementary propagator (Wick) contractions of the four points
\bea
{\cal X}=\frac{(12)(34)}{x_{12}^2x_{34}^2} \, , ~~~~
{\cal Y}=\frac{(13)(24)}{x_{13}^2x_{24}^2} \, , ~~~~
{\cal Z}=\frac{(14)(23)}{x_{14}^2x_{23}^2} \, . \label{XYZ}
\eea
Here $(12) =(21) \equiv u_1^au_2^a$, etc. denote the SO(6) invariant contractions of the harmonic variables at two different points. Every monomial
in eq. (\ref{gampl}) corresponds to a certain propagator structure built out of the propagators of the elementary fields (see Fig. 1). We refer to eq. (\ref{gampl}) as to the representation of the amplitude in the propagator basis. The coefficients $a_{mnl}^{(k)}(s,t)$ are
arbitrary functions of the two independent conformal invariant cross-ratios
$$
s=\frac{x^2_{12}\, x^2_{34}}{x^2_{13}\, x^2_{24}}\,, \qquad
t=\frac{x^2_{14}\, x^2_{23}}{x^2_{13}\, x^2_{24}}\,.
$$

Superconformal symmetry puts additional kinematical restrictions on the coefficient functions in (\ref{gampl}) which take the form of differential equations (superconformal Ward identities) \cite{Eden:2000qp,DO}. Further, dynamical constraints on the amplitude are obtained through the procedure of inserting the ${\cal N}= 2$ SYM action. According to the ``partial non-renormalization theorem" of Ref. \cite{EPSS}, the quantum ${\cal N}= 2$ amplitude takes a {\it factorized form} (see  Section 4 for details). Recently this insertion procedure has been generalized to the ${\cal N}= 4$ case \cite{ADOS,AS}.  The factorized form of the quantum corrections to the ${\cal N}=4$ amplitude is
\bea
\label{N4ins} \la k|k|k|k \ra^{\rm\tiny quant.}=R_{{\cal
N}=4}\sum_{m+n+l=k-2}{\cal F}_{mnl}^{(k)}(s,t) {\cal X}^m{\cal
Y}^n{\cal Z}^l \, , \label{Intform} \eea where $R_{{\cal
N}=4}$ is a {\it universal polynomial prefactor} carrying weight 2 at each point.
Explicitly,
\begin{eqnarray}
  R_{{\cal N}=4}&\ea & s\ {\cal X}^2 +
 {\cal Y}^2 +
t \  {\cal Z}^2  +
   (s-t-1)\ {\cal Y}{\cal Z}
\nonumber\\
  &&
 {}  +  (1-s-t)\ {\cal X}{\cal Z}
 +  (t-s-1) \ {\cal X}{\cal Y}\, . \label{R}
\end{eqnarray}
All the dynamical information is thus encoded in the
conformally invariant coefficient functions ${\cal
F}_{mnl}^{(k)}(s,t)$ which  are the subject of our
subsequent investigation.\footnote{To make the reader familiar with
the notation we stress that the upper index in ${\cal
F}_{mnl}^{(k)}(s,t)$ denotes the weight of the $\half$-BPS operators. For the
sake of clarity we sometimes separate the lower indices by
commas.} Substituting the universal prefactor into eq.
(\ref{Intform}) and comparing it with eq. (\ref{gampl}), we read
off the following expression for the {\it quantum corrections} to the coefficients
$a_{mnl}^{(k)}$: \bea \label{aF}
a_{mnl}^{(k)}(s,t)&=&s{\cal F}_{m-2,n,l}^{(k)}+{\cal F}_{m,n-2,l}^{(k)}+t{\cal F}_{m,n,l-2}^{(k)}\\
\nonumber
&+&
 (s-t-1)\ {\cal F}_{m,n-1,l-1}^{(k)}+(1-s-t)\  {\cal F}_{m-1,n,l-1}^{(k)}
+ (t-s-1) \  {\cal F}_{m-1,n-1,l}^{(k)} \, .
\eea
If any of the indices of ${\cal F}_{mnl}^{(k)}$ in this equation becomes 
negative, then the corresponding term is absent.

Since all the operators involved are identical, the four-point amplitude is invariant
under the symmetric group $S_4$ which permutes the points  $1,\ldots,4$.
Only the subgroup $S_3$ acts non-trivially on the cross-ratios $s$ and $t$ and, as a result, on the coefficient functions ${\cal F}_{mnl}^{(k)}$.
It is generated by the two independent crossing transformations, e.g., $1\leftrightarrow 2$
and $1\leftrightarrow 3$.
If all the indices
$m,n,l$ of ${\cal F}_{mnl}^{(k)}$ are different, one can use the action of $S_3$
to order them, e.g., $m> n > l$. The function  ${\cal F}_{mnl}^{(k)}$ with $m> n > l$
does not satisfy any crossing symmetry relation. It can be taken as a representative of the corresponding crossing equivalence class consisting of six elements. If any two indices of ${\cal F}_{mnl}^{(k)}$ coincide, then the function satisfies an additional crossing symmetry relation (and the corresponding crossing equivalence class consists of three elements). Finally, in the case where all three indices coincide, the function transforms into itself under the whole
group $S_3$ (and the corresponding crossing equivalence class consists of a single element).

\vskip 15pt
\noindent
\begin{minipage}{\textwidth}
\begin{center}
\includegraphics[width=0.40\textwidth]{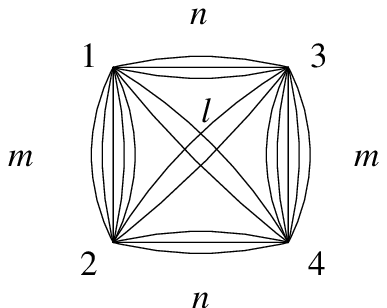}
\end{center}
\begin{center}
Figure 1. The propagator structure corresponding to the monomial 
${\cal X}^m{\cal Y}^n{\cal Z}^l$. 
\end{center}
\end{minipage}
\vskip 20pt

For sufficiently small values of $k$ the number of conformal invariants $\cF$
representing the independent crossing equivalence classes is easy to work out (the general formula is given in Ref. \cite{HH}). For instance, both for $k=2,3$ the quantum part of the amplitude is given by a single function $\cF$, while already for $k=4$ two functions are needed. 

The functions $\cF$ can be explicitly computed both in
perturbation theory and in the strong coupling regime by using
the AdS/CFT duality. In Ref. \cite{AS} it was observed that in the
large $N$ limit the perturbative four-point amplitudes of the
$\half$-BPS operators exhibit a certain {\it degenerate behavior}
in comparison to their supergravity partners: The number of
independent conformal invariants describing the one-loop (order $\lambda \sim g^2$) perturbative amplitude is smaller than that in the supergravity regime. In what follows we  study this degeneracy phenomenon at the two-loop (order $\lambda^2 \sim g^4$) level by explicitly working out the
corresponding four-point amplitude.

Computing four-point correlation functions at two loops
is of course much more involved than at one loop, due to the rapidly increasing number
of Feynman graphs. A very efficient technique is provided by the
${\cal N}=2$ harmonic superspace formalism. This approach, combined with
the ${\cal N}=2$ insertion procedure, allows us to drastically reduce the number of relevant graph topologies to be computed.


In order to apply the ${\cal N}=2$ harmonic superspace formalism we need
to decompose the $\half$-BPS operators into their ${\cal N}=2$ hypermultiplet (HM) and SYM constituents. This is done by first decomposing the ${\cal N}=4$ field-strength superfield ${\cal W} = u^a\phi^a(x) + \cdots$ into the Grassmann analytic HM superfield $q^+ = u^+_i \phi^i(x) + \cdots$ (and its harmonic conjugate\footnote{Here $\widetilde{q}$ means the usual complex conjugation for the field together with an antipodal reflection on the sphere $S^2\sim $ SU(2)/U(1) for the harmonic variable.} $\widetilde q^+ = u^+_i \bar\phi^i(x) + \cdots$) and the chiral field strength $W = w(x) + \cdots$ (and its antichiral conjugate $\bar W = \bar w(x) + \cdots$). Here the six real scalars $\phi^a(x)$ forming an SO(6) vector have been split into an SU(2) doublet $\phi^i(x)$, $i=1,2$ and a complex singlet $w(x)$. The SO(6) vector harmonics $u^a$ have been replaced by SU(2) fundamental harmonics $u^\pm_i$ ($u^- = (u^+)^*$, $u^+_i\epsilon^{ij}u^-_j = 1$), parametrizing the sphere $S^2\sim $ SU(2)/U(1). 

In the sequel we will be interested in the ``pure'' HM projection in which we put only $\tq$'s at points 1 and 4 and only $q$'s at points 2 and 3.
As explained in Ref. \cite{AS}, taking this projection means considering only the terms with $l=0$ in eq. (\ref{gampl}) (in Fig. 1 they correspond to the graphs without diagonals). Further, the propagator structures (\ref{XYZ}) are replaced by \bea
{\cal X}\to X=\frac{[12][43]}{x_{12}^2x_{34}^2} \, , ~~~~
{\cal Y}\to Y=\frac{[13][42]}{x_{13}^2x_{24}^2} \, ,
\eea
where, e.g., 
\begin{equation}\label{harcontra}
  [12] =-[21] \equiv (u_1)_i^+ \epsilon^{ij} (u_2)_j^+
\end{equation}
is the SU(2) invariant contraction of the two sets of harmonic variables at points 1 and 2. Under this projection eq. (\ref{N4ins}) reduces to
\bea
\label{pp}
\la Q^{(k)}\ra\equiv \la \tq^k|q^k|q^k|\tq^k\ra= R_{{\cal N}=2}
\sum_{m=0}^{k-2}\cF_{m,k-m-2,0}^{(k)}(s,t)\ X^m Y^{k-m-2} \, ,
\eea
where
\bea
\label{RN2}
R_{{\cal N}=2}=
sX^2+(t-s-1)\ XY+Y^2 \, .
\eea

The factorized expression (\ref{pp}) with the universal prefactor (\ref{RN2}) is precisely what one finds by the direct application of the field-theoretic insertion procedure to the four-point amplitude in the ${\cal N}=2$ theory (see Section 4).

According to the previous discussion of the crossing symmetry, we can take
$\cF_{m,k-m-2,0}^{(k)}$ with
\bea
\label{meven}
m_{-} \equiv 1/2(k-2)\leq m\leq  k-2\equiv m_{+}~~~~~~\mbox{for $k$ even}
\eea
and
\bea
\label{modd}
 1/2(k-1)\leq m\leq  k-2\equiv m_{+}~~~~~~\mbox{for $k$ odd}
\eea
as representatives
of the different crossing equivalence classes of the coefficient
functions of the four-point amplitude. 

It should be pointed out that for
$k>4$ the whole ${\cal N}=4$ amplitude cannot be restored from its
pure ${\cal N}=2$ projection\footnote{To this end we would also need some of the ``mixed" projections 
of the type $\tq^{k-p}q^p$. Obtaining such projections involves some 
amount of linear algebra; in addition, the corresponding two-loop graphs 
are more complicated. Therefore here we prefer to restrict ourselves to 
the ``pure" sector.} and consequently, the functions
$\cF_{m,k-m-2,0}^{(k)}$ above
do not form a complete set of {\it a priori} independent conformal
invariants needed to parametrize the amplitude. However, the subset of
conformal invariants emerging in the pure HM projection
is sufficiently large to illustrate the large $N$ degeneracy problem
at the two-loop level.

Among the chosen set of independent coefficient functions only one 
(for $k$ odd)
or two (for $k$ even) admit a pair of coincident indices. 
The first such function  ${\cal F}^{(k)}_{+}(s,t)\equiv \cF_{k-2,0,0}(s,t)$ 
is obtained by setting $m=m_{+}$, so it exists for any value of $k$.\footnote{One
could instead take $\cF_{0,k-2,0}^{(k)}$
 which obeys
$\cF_{0,k-2,0}^{(k)}(s,t) =\cF_{0,k-2,0}^{(k)}(t,s)$. The
relation between the two functions is
$\cF_{k-2,0,0}^{(k)}(s,t)=1/s~\cF_{0,k-2,0}^{(k)}(1/s,t/s)$.}
It has the additional symmetry property 
\bea
\cF^{(k)}_{+}(s,t)=1/t~\cF^{(k)}_{+}(s/t,1/t) \, . \eea 
The second function $\cF_{m_{-},m_{-},0}^{(k)}$ is obtained by setting $m=m_{-}$, so it exists only for $k$ even. In this case we have \bea
\cF_{m_{-},m_{-},0}^{(k)}(s,t)=1/s~\cF_{m_{-},m_{-},0}^{(k)}(1/s,t/s)
\, . \eea
It is more convenient to introduce the functions 
\bea \label{Newparam} {\cal
F}^{(k)}_m(s,t)\equiv \cF_{m,k-m-2,0}^{(k)}(t,s)\, ,~~~~m\neq m_+ 
\eea 
since for $m=m_{-}$ the function ${\cal F}^{(k)}_-(s,t) \equiv 
\cF_{m_{-},m_{-},0}^{(k)}(t,s)$ has the same crossing symmetry as 
${\cal F}^{(k)}_{+}$: 
\bea
\label{pmcross} \cF^{(k)}_{\pm}(s,t)=1/t~\cF^{(k)}_{\pm}(s/t,1/t)
\, . \eea 
The functions $\cF^{(k)}_m$ with $m_{-}< m < m_{+}$ {\it do not a priori obey any crossing symmetry relations}. Finally, we note that the case $k=2$ is exceptional as
all the three indices of ${\cal F}^{(2)}$ vanish.
This function transforms covariantly under the whole
group $S_3$.

In summary, the pure HM projection is parametrized by the chain of conformally
invariant functions $\cF^{(k)}_m$ whose ``boundaries'' $\cF^{(k)}_{\pm}$
are subject to the crossing symmetry relation (\ref{pmcross}). The function
$\cF^{(k)}_{-}$ exists only for $k$ even.


Concluding this section, we give the main result of our two-loop calculation, 
the details of which are presented in Sections 4 and 5. 
In  the large $N$ limit we find
\bea
\nonumber &&\cF^{(k)}_{-}(s,t)=\frac{\lambda^2}{N^2}\frac{k^2}{4}
\Big\{
\frac{1}{4}s[\Phi^{(1)}(s,t)]^2+\frac{1}{s}\Phi^{(2)}(t/s,1/s)
\Big\}=\cF^{(k)}_m(s,t)\, , \\
\label{minpl}
&&\cF^{(k)}_{+}(s,t)=\frac{\lambda^2}{N^2}\frac{k^2}{4}\Big\{ \frac{1}{4}(t+1)
[\Phi^{(1)}(s,t)]^2
+\Phi^{(2)}(s,t)
+\frac{1}{t}\Phi^{(2)}(s/t,1/t)
\Big\}\, ,
\eea
where the range of  $m$ is given by Eqs. (\ref{meven}) and 
(\ref{modd}), in particular, for $k$ odd the function $\cF^{(k)}_{-}$ is absent.
In eqs. (\ref{minpl}) $\lambda= {g^2N}/{4\pi^2}$ is the 't Hooft coupling and the functions 
$\Phi^{(1)}$ and $\Phi^{(2)}$ are the so-called one- and two-loop scalar box 
conformal integrals
\bea
\label{h1}
&&\int\frac{\rmd^4 x_5}{x_{15}^2x_{25}^2x_{35}^2x_{45}^2}\equiv 
-\frac{i\pi^2}{x_{13}^2x_{24}^2} \,
\Phi^{(1)}(s,t) \, , \\
\label{h2}
&& \int\frac{x_{13}^2 ~\rmd^4 x_5\rmd^4 x_6}{x_{15}^2x_{25}^2x_{35}^2x_{56}^2x_{16}^2x_{36}^2x_{46}^2}\equiv \frac{(i\pi^2)^2}{x_{13}^2x_{24}^2}\, \Phi^{(2)}(s,t) \, .
\eea

We see that {\it in the large $N$ limit all but one conformal invariants 
describing the pure HM projection are equal}. It is only the ``boundary'' 
function $\cF_{+}^{(k)}$ which
is distinctly different from the others.

Having found the pure HM part of the four-point amplitude of the
$\half$-BPS operators, we can now understand what happens to the large
$N$ degeneracy at the two-loop level. According to Ref. \cite{AS},
at one loop and in the large $N$ limit all conformal invariants
emerging in the pure HM projection coincide: 
\bea
\cF_{\pm}^{(k)}(s,t)=\cF_m^{(k)}(s,t)
=-\frac{\lambda}{N^2}\frac{~k^2}{2}~\Phi^{(1)}(s,t)
\, . 
\eea 
This degeneracy manifests itself already in the case $k=3$: 
The single function,
$\cF_+^{(3)}$, describing the one-loop amplitude obeys the extra crossing
symmetry relation $\cF_+^{(3)}(s,t)=\cF_+^{(3)}(t,s)$  
which is not required on general grounds. Just as in the one-loop case, 
the two-loop function $\cF_+^{(3)}$ from eq. (\ref{minpl})
obeys eq. (\ref{pmcross}) imposed by the symmetry of the amplitude, but 
the one-loop ``bonus'' crossing symmetry relation 
does not hold anymore. We therefore conclude that this symmetry of the 
$k=3$ one-loop amplitude is destroyed by the two-loop quantum corrections.

Another type of degeneracy appears in the case $k=4$. On the general grounds 
of superconformal and crossing symmetry two {\it a priori} independent 
conformal invariants are needed in this case, while at one loop
both of them appear to coincide in the large $N$ limit \cite{AS}.
According to eqs. (\ref{minpl}), at the two-loop level we indeed
find two conformally invariant functions $\cF^{(4)}_{\pm}$  which
are distinctly different. Again we observe that the one-loop degeneracy is 
lifted when the two-loop corrections are switched on.

However, we also see that both types of large $N$ degeneracy mentioned above
{\it still hold at the two-loop level for operators of higher weight, $k>4$}. 
Indeed, for $k>4$ the functions $\cF_m^{(k)}$ coincide (they are equal to 
$\cF^{(k)}_-$ for $k$ odd) and have the same bonus crossing symmetry 
(\ref{pmcross}) as $\cF^{(k)}_-$.

Finally, we mention that in the special case $k=2$ we find
\bea
\cF^{(2)}(s,t)&=&\frac{\lambda^2}{N^2}
\Big[ \frac{1}{4}(s+t+1) [\Phi^{(1)}(s,t)]^2\\
\nonumber
&&~~~~~~+\frac{1}{s}\Phi^{(2)}(t/s,1/s)
+
\Phi^{(2)}(s,t)+\frac{1}{t}\Phi^{(2)}(s/t,1/t)
\Big]\, .
\eea
This result has been previously obtained in Refs. \cite{ESS,BKRS}.

\section{Mixing and large $N$ degeneracy}

In field theory $\half$-BPS operators can be realized as mixtures of
single- and multi-trace operators.
Here we investigate whether the operator mixing can affect the degeneracy problem of four-point correlation functions. 
As an explicit example
we consider the case of weight $k=4$ operators.

As discussed in Ref. \cite{AS}, for $\half$-BPS operators of weight $k\leq 4$
the complete
${\cal N}=4$ four-point amplitude can be reconstructed from the ``pure''
${\cal N}=2$ hypermultiplet
projection, eq. (\ref{pp}). In particular, in the case $k=4$ we can write down
\bea
\langle Q^{(4)} \rangle &=&
\sum_{m=0}^4 a_{m,4-m,0}^{(4)}(s,t) X^mY^{4-m}\, .
\label{1'}
\eea
This projection involves five of the fifteen coefficient functions of the original
${\cal N}=4$ amplitude, which belong to the three crossing equivalence
classes
$(a_{040}^{(4)},a_{400}^{(4)})$, $(a_{130}^{(4)},a_{310}^{(4)})$ and
$a_{220}^{(4)}$.

In order to study possible mixing effects on $Q^{(4)}$ we compute
the coefficients $a^{(4)}$ at tree level. There, generically,
the $\half$-BPS operator is of the form
\begin{equation}
{\cal O} \sim S + a D \, ,
\label{mixed}
\end{equation}
where we have defined
\begin{equation}
S \equiv {\rm Tr}(q^4) \, , \qquad \quad D 
\equiv {\rm Tr}(q^2){\rm Tr}(q^2)\,,
\end{equation}
$q$ being the ${\cal N}=2$ Grassmann analytic HM superfield and $a$ an
$N$-dependent mixing parameter.

We fix the normalization coefficient for ${\cal O}$ by requiring
its two-point function to be normalized canonically.
Computing separately the two--point functions $\langle SS \rangle$,
$\langle SD \rangle$ and $\langle DD \rangle$, in the case of SU(N)
color group we find 
(since we are interested in the color combinatorics we do not
exhibit the space-time dependence of the correlators)
\bea
&& \langle SS \rangle =~ 4 (N^2-1) ~{\cal P}_1\, ,
\non\\
\non
&& \langle SD \rangle =~ 8(N^2-1) ~{\cal P}_2 \, ,
\\
\label{2pt}
&& \langle DD \rangle =~  8(N^2-1) ~{\cal P}_3 \, ,
\eea
where, for convenience, we have introduced
\bea
 {\cal P}_1 ~=~ \frac{1}{N^2}(N^4-6N^2+18)\, , ~~~~
 {\cal P}_2 ~=~ \frac{1}{N}(2N^2-3) \, , ~~~~
 {\cal P}_3 ~=~ (N^2+1) \, .
\label{pol}
\eea
Therefore, the canonically normalized operator reads
\begin{equation}
{\cal O} = \Big[4 (N^2-1) \left( {\cal P}_1 + 4a
{\cal P}_2 +2a^2 {\cal P}_3 \right) \Big]^{-\frac12} (S + aD)\,.
\label{normalized}
\end{equation}

The tree level contributions to the coefficients $a^{(4)}$ are given by
diagrams with propagator structures as in Fig. 1, with $l=0$, $n=4-m$,
$m=0, \cdots, 4$.
The coefficient functions $a_{040}^{(4)}$ and $a_{400}^{(4)}$ correspond to
disconnected diagrams built as product of 2-point function diagrams.
Therefore, given the above normalization, they are trivially equal to $1$.
In order to determine the other coefficients we need to compute the actual
four-point amplitudes for the mixed operator ${\cal O}$. This can be achieved 
by evaluating separately the four-point functions
$\langle SSSS \rangle$, $\langle DDDD \rangle$ and $\langle SSSD \rangle$,
$\langle SSDD\rangle$,
$\langle SDDD\rangle $ as well as the amplitudes obtained from the latter
by permuting $S$ and $D$.

Since it is enough to compute one coefficient function within each crossing
equivalence class, we concentrate, for instance, on the coefficient
function $a_{310}^{(4)}$.
It is convenient to first evaluate the building blocks of Fig. 2.

\vskip 15pt
\noindent
\begin{minipage}{\textwidth}
\begin{center}
\includegraphics[width=0.60\textwidth]{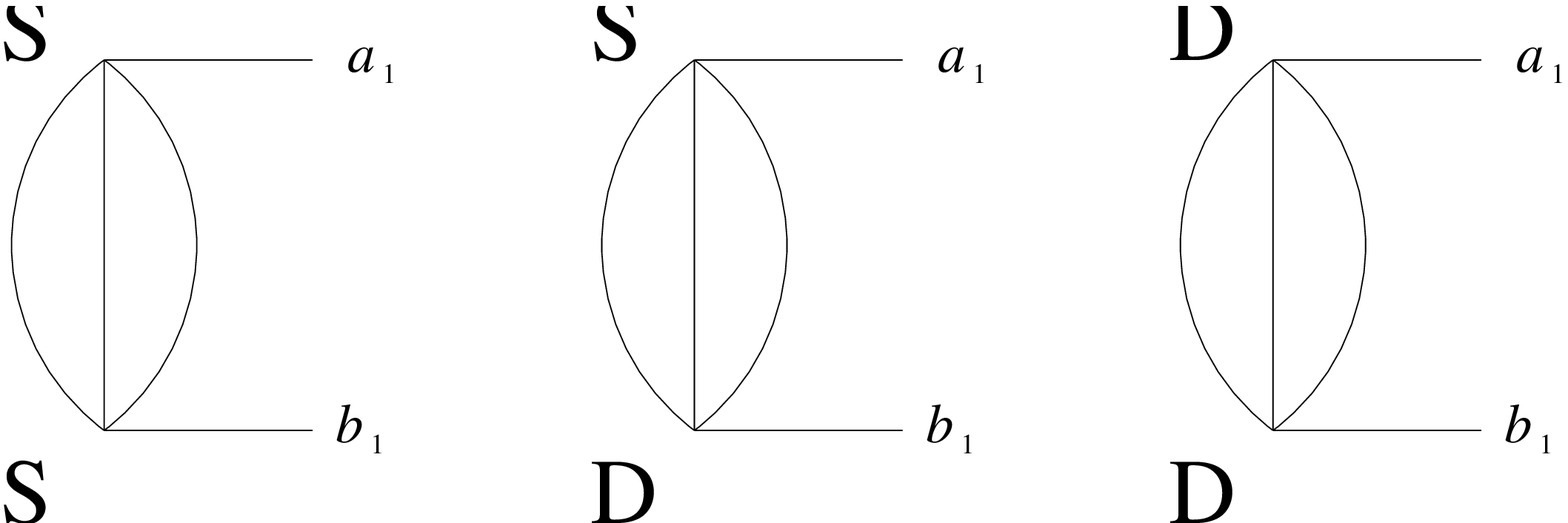}
\end{center}
\begin{center}
Figure 2. Building blocks for the coefficient $a_{310}^{(4)}$.
\end{center}
\end{minipage}
\vskip 20pt

These building blocks can be obtained from the tree diagrams of the
2-point amplitudes (\ref{2pt}) by cutting a line. 
This operation simply amounts to
inserting a $\d_{a_1 b_1}$ for the line cut without affecting the color
structure. Therefore, the three building blocks are still proportional
to the polynomials given in eq. (\ref{pol}). Explicitly,
\bea
&& SS \rightarrow 2^4 {\cal P}_1 ~\d_{a_1 b_1}
\nonumber \\
&& SS \rightarrow 2^5 {\cal P}_2 ~\d_{a_1 b_1}
\nonumber \\
&& SS \rightarrow 2^5 {\cal P}_3 ~\d_{a_1 b_1}
\label{pol2}
\ena
These building blocks are then combined to realize
all possible configurations of single- and double-trace operators. 
The result is
\bea
a_{310}^{(4)}=a_{130}^{(4)}&=& \frac{2^8(N^2-1)\left( {\cal P}_1^2
+ 16 a^2 {\cal P}_2^2 + 4a^4 {\cal P}_3^2 + 8a {\cal P}_1 {\cal P}_2
+ 16a^3 {\cal P}_2{\cal P}_3 + 4a^2 {\cal P}_1{\cal P}_3
\right)}{2^4 (N^2-1)^2 \left( {\cal P}_1 + 4a
{\cal P}_2 +2a^2 {\cal P}_3 \right)^2}
\non\\
& =& \frac{2^8 (N^2-1)
\left( {\cal P}_1 + 4a {\cal P}_2 +2a^2 {\cal P}_3
\right)^2}{2^4 (N^2-1)^2 \left( {\cal P}_1 + 4a
{\cal P}_2 +2a^2 {\cal P}_3 \right)^2}
= \frac{16}{N^2-1} \, ,
\label{b1free}
\eea
Being independent of the mixing parameter $a$, it gives the
finite $N$ result for the tree level part of the coefficient
function $a_{310}^{(4)}$, independently of the presence of the double-trace 
operator.
In the large $N$ limit this expression agrees with the result
of Ref. \cite{AS}.
We note that the independence of $a$ is due to the particular structure
(\ref{pol2})
of the building blocks. For $k\geq 4$ we observe the same phenomenon in the 
coefficients $a_{(k-1)10}^{(k)}$.

Next we concentrate on $a_{220}^{(4)}$.  Again, the calculation can
be carried out by first computing the building blocks depicted in Fig. 3.

\vskip 15pt
\noindent
\begin{minipage}{\textwidth}
\begin{center}
\includegraphics[width=0.60\textwidth]{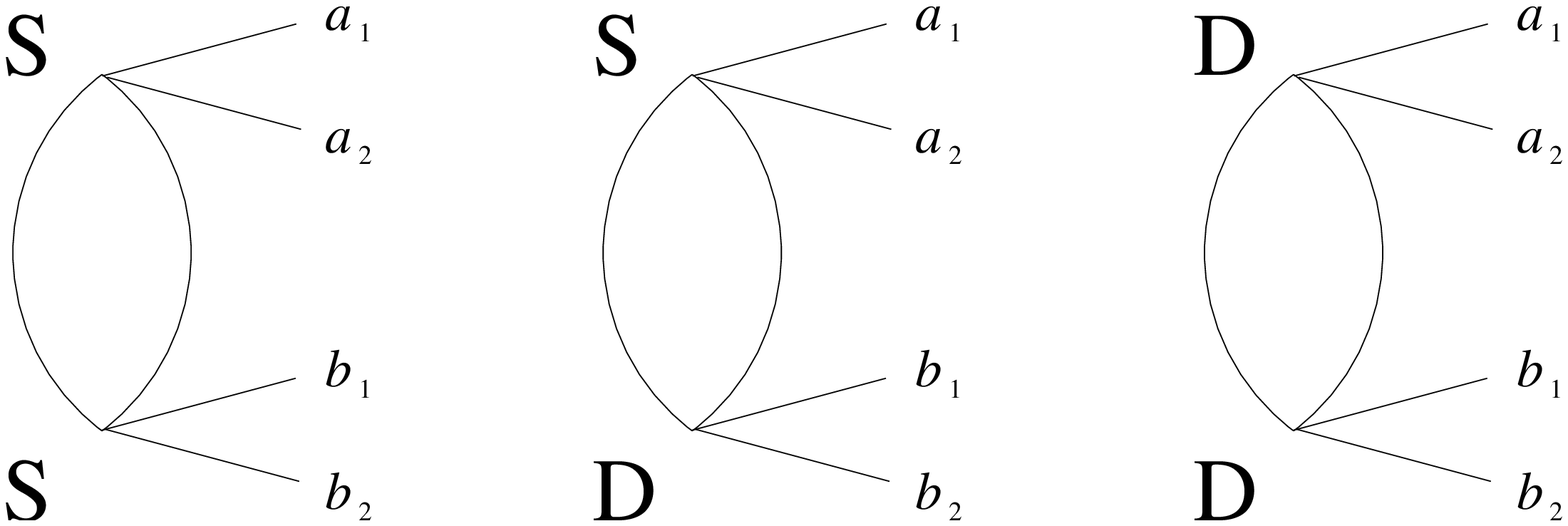}
\end{center}
\begin{center}
Figure 3. Building blocks for the coefficient $a_{220}^{(4)}$.
\end{center}
\end{minipage}

\vskip 20pt
Introducing the concise notation ${\rm Tr}(T_aT_b...T_cT_d) \equiv (ab...cd)$
for the trace of the color generators,
the algebraic expressions for the building blocks of Fig. 3 can be written as
\bea
&& SS ~\rightarrow~ 4 \frac{(N^2-9)}{N} \left[ (a_1 a_2 b_1b_2) +
(a_2 a_1 b_1b_2)+ (a_1 a_2 b_2 b_1)+ (a_2 a_1 b_2 b_1) \right]
\non\\
&&~~~~~~~~~~~+ 8\frac{2N^2+9}{N^2} (a_1a_2)(b_1b_2) +
4 (a_1b_1)(a_2b_2) + 4 (a_1b_2)(a_2b_1) \, ;
\non\\
&& SD ~\rightarrow~ 2^3 \Big[ (a_1 a_2 b_1b_2) +
(a_2 a_1 b_1b_2)+ (a_1 a_2 b_2 b_1)+ (a_2 a_1 b_2 b_1) + (a_1b_1a_2b_2)
+ (a_1b_2a_2b_1)
\non\\
&&~~~~~~~~~~~ + \frac{(2N^2-3)}{N} ~(a_1a_2)(b_1b_2) \Big] \, ;
\non\\
&& DD ~\rightarrow~ 2^3 \left[ (N^2+3) (a_1a_2)(b_1b_2) +
2(a_1b_1)(a_2b_2) + 2(a_1b_2)(a_2b_1) \right] \, .
\non
\eea
All possible configurations of the four-point amplitudes of $S$ and $D$
are obtained by combining the building blocks of
Fig. 3 in all possible ways. The last contraction between two blocks 
produces an extra factor of $4$. 
Pulling out the common factor $2^8 (N^2-1)$, we find
\bea
&& \langle SSSS \rangle ~\rightarrow~ \left(N^4 - \frac{15}{2} N^2 +135 -
\frac{729}{N^2}
+ \frac{1215}{N^4}\right) ~\equiv {\cal P}_4 \, ,
\non\\
&& \langle SSSD \rangle
~\rightarrow~ \frac{1}{N^3}(10 N^6 - 54N^4 +216 N^2 -405)
~\equiv {\cal P}_5 \, ,
\non\\
&& \langle SSDD \rangle ~\rightarrow~ \frac{1}{N^2}(4N^6 + 6N^4 -63N^2 +135)
~\equiv {\cal P}_6 \, ,
\non\\
&& \langle SDDD \rangle ~\rightarrow~ \frac{1}{N}(2N^6 + 5N^4 +18N^2 -45)
~\equiv {\cal P}_7 \, ,
\non\\
&& \langle DDDD \rangle ~\rightarrow~ (N^6 + 5N^4 +19N^2 +15)
~\equiv {\cal P}_8
\label{pol3}
\eea
The final answer for the coefficient $a^{(4)}_{220}$ is then
\begin{equation}
a_{220}^{(4)} ~=~ \frac{16 \left({\cal P}_4 + 4a {\cal P}_5 + 6a^2 {\cal P}_6
+ 4a^3 {\cal P}_7 + a^4 {\cal P}_8 \right)}{(N^2-1)
({\cal P}_1 + 4a {\cal P}_2 + 2a^2 {\cal P}_3)^2}
\label{c1freemixed}
\end{equation}
We notice that by setting $a=0$ (i.e., by neglecting the double--trace
contributions) we obtain the finite $N$ result for the free amplitude
involving only the single--trace operator:
\begin{equation}
a_{220}^{(4)~\rm \tiny single} ~=~ \frac{16(N^4 - \frac{15}{2} N^2 +135 - \frac{729}{N^2}
+ \frac{1215}{N^4})}{(N^2-1)(N^2 -6+ \frac{18}{N^2})^2}
\label{c1free}
\end{equation}
In the large $N$ limit this expression agrees with that in Ref. \cite{AS}.

Next we analyze the large $N$ limit of the general expression
(\ref{c1freemixed}).
Taking the large $N$ limit of ${\cal P}_i$ we obtain
\begin{equation}
a_{220}^{(4)} ~=~ \frac{16 \left(N^2 + 40(N a)  + 24(Na)^2
+ 8(Na)^3  + (Na)^4  \right)}{
(N^2+ 8(Na)  + 2(Na)^2)^2} \, .
\label{c1freemixed1}
\end{equation}
If in the large $N$ limit the mixing parameter is chosen to be
of order one, then the
double-trace part of $\cal O$ provides a dominant contribution and one finds
$a_{220}^{(4)}\sim 1$.

We recall that according to our general treatment in Section 2 
the coefficient function $a_{220}^{(4)}(s,t)$ admits a unique 
splitting on the non-trivial (``quantum'') function of the conformal cross-ratios
constructed out of $\cF_{\pm}$
and an integration (``free field'') constant $a_{220}^{(4)}$
\bea
a_{220}^{(4)}(s,t)=a_{220}^{(4)}+\cF_+(s,t)+\cF_+(1/s,t/s)+(t-s-1)\cF_-(t,s) \, .
\eea
The constant part of $a_{220}^{(4)}(s,t)$ was 
computed in the supergravity regime \cite{AS} with the result 
$a_{220}^{(4)~\rm \tiny sugra}=\frac{16}{N^2}$, i.e. it scales as $1/N^2$.
Therefore, the AdS/CFT correspondence does not allow the mixing parameter $a$ to be of order one.
If we pick $a=\kappa/\sqrt{N}+\ldots$, then it
will affect the coefficient $a_{220}^{(4)}$ at order $1/N^2$:
\bea
a_{220}^{(4)}=\frac{16}{N^2}(1+\kappa^4) \, .
\eea
However, if we want $a_{220}^{(4)}$ to match the large $N$ value
computed from supergravity exactly,
we ought to require $\kappa=0$, i.e. the mixing parameter has to scale
faster than $1/\sqrt{N}$.
Moreover, we note that if $a$ is expanded in the odd powers of $1/N$, then the free amplitude has a 't Hooft
type expansion in powers of $1/N^2$. Therefore, the simplest assumption for
the mixing coefficient is $a \sim 1/N+\ldots $.

\vskip 15pt

\noindent
\begin{minipage}{\textwidth}
\begin{center}
\includegraphics[width=0.45\textwidth]{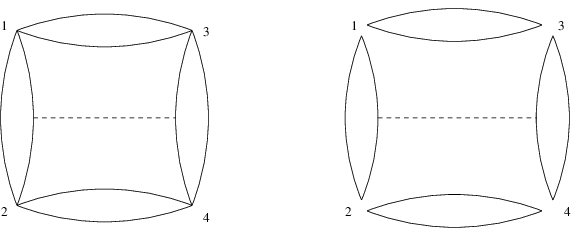}
\end{center}
\begin{center}
Figure 4. The leading large $N$ contributions to the four-point amplitudes
$\langle SSSS\rangle $
and $\langle DDDD\rangle $ at order $\lambda$.
\end{center}
\end{minipage}
\vskip 20pt

Let us now briefly discuss the influence of the mixing on the large $N$
degeneracy occurring at the one-loop level (order $\lambda \equiv
\frac{g^2}{4\pi}$).
As shown in Ref. \cite{AS}, harmonic analyticity, i.e. the superconformal
kinematics, allows one to reduce the one-loop calculation
of the four-point amplitude just to a single Feynman graph given in Fig. 4.
This statement holds irrespectively of the intrinsic trace structure of the
BPS operators
and therefore it can be applied to the operator ${\cal O}$.

The various contributions to the coefficient $a_{220}^{(4)}$
from the amplitude at order $\lambda$
are then obtained by specifying the operator ($S$ or $D$) at each corner
of the graph in Fig. 4. Using a double line notation for the propagators
in there, the leading large $N$ behavior is easily identified:
Both the purely single-trace part $\langle SSSS\rangle $ and the
purely double-trace part $\langle DDDD\rangle$ grow as $N^7$.
This is in contradistinction to the free field case, where
$\langle SSSS\rangle \sim N^6$ while $\langle DDDD\rangle \sim N^8$
(see eq. (\ref{pol3})).
Therefore, at one loop the contribution of the double-trace operator appears
to be more suppressed than in the free case, if compared to that of the single trace.
The double-trace  contribution can compete with the single-trace one if the mixing parameter is of order one. This is however ruled out by the free field considerations
above which set $a$ to scale faster than $1/\sqrt{N}$.

In conclusion, we see that if one requires
the large $N$ free field theory amplitude to coincide with the one for
single-trace operators
(and therefore with the constant part of the supergravity induced amplitude),
then
the mixing produces no effect on the leading large $N$ one-loop amplitude
and, as a consequence, on its degeneracy.
Such a behavior follows the supergravity pattern:
The redefinitions of the supergravity fields (corresponding to the field-theoretic mixing)
produce no effect on the four-point amplitude, provided that it is of the regular
(i.e., not of the extremal or sub-extremal) type \cite{AF}.

\section{The ${\cal N}=2$ insertion procedure and Feynman rules}

We would like to find the quantum corrections to the lowest component (at $\theta_{1,2,3,4}=0$)
of the four-point correlator $\la Q^k \ra$ and identify the corresponding conformal
invariants $\cF$ in eq. (\ref{pp}). Already in the one-loop case, but even more so at two loops, the most efficient
technique, in our opinion, is the insertion procedure in  ${\cal N}=2$ harmonic superspace \cite{Howe:2000hz,EPSS,Eden:2000mv,ESSun}.

The basic ingredients of the  ${\cal N}=4$ SYM theory in terms of ${\cal N}=2$ superfields are the hypermultiplet $q^+(x,\theta^+,\bar\theta^+,u)$ and the ${\cal N}=2$ SYM field strength $W(x,\theta)$. The distinctive feature of the former is that it is a Grassmann analytic superfield (i.e., it depends only on the harmonic U(1) projections $\theta^+,\bar\theta^+$) while the latter is a chiral superfield.

One of the advantages of the  ${\cal N}=2$ formulation of the theory compared to  ${\cal N}=1$ is that the HM composite operators like ${\rm Tr} (q^+)^k$ need no covariantization (no presence of the gauge superfield at the external points of the amplitude). Further, the HM matter interacts with the gauge sector only through a single cubic vertex. The true non-Abelian nature of the theory is encoded in the gauge self-interactions (as well as in the ghost sector, but we do not see it at the level $g^4$). 

The quantum corrections to the four-point correlator $\la Q^k \ra$ at order
$g^4$ can be obtained by a double insertion of the ${\cal N}=2$ SYM action
\bea 
S_{{\cal N}=2~\mbox{\tiny SYM}}=\int \rmd^4
x\rmd^4\theta~ {\cal L}\, ,
~~~~{\cal L}=\frac{1}{4g^2}\mbox{Tr}W^2 \, .
\eea 
More precisely,
\bea 
\label{ins} \langle Q^{(k)} \rangle_{g^4} = \frac{1}{2} \int \rmd^4
x_5\rmd^4\theta_5\ \rmd^4
x_6\rmd^4\theta_6 ~\langle Q^{(k)} |{\cal L}(5)|{\cal L}(6) \rangle_{\rm\tiny tree}\, .
\eea 
Here $\langle Q^{(k)} |{\cal L}(5)|{\cal L}(6) \rangle_{\rm\tiny tree}$ is a new, six-point correlator calculated at ``tree level" (Born approximation). The corresponding Feynman graphs are obtained by drawing the usual two-loop (order $g^4$) graphs for $\langle Q^{(k)} \rangle$ and then inserting the linearized ${\cal N}=2$ SYM Lagrangian ${\cal L}$ into the gluon lines twice, each time into a different gluon line.\footnote{This insertion procedure is based on the formula for the second derivative of the four-point correlator with respect to the gauge coupling
constant: $\left({\pa}/{\pa
g^2}\right)^2\langle Q^{(k)} \rangle= \frac{2}{g^4} \int \rmd^4
x_5\rmd^4\theta_5\ ~\langle Q^{(k)} |{\cal L} \rangle +\frac{1}{g^4} \int \rmd^4
x_5\rmd^4\theta_5\ \rmd^4
x_6\rmd^4\theta_6 ~\langle Q^{(k)} |{\cal L}|{\cal L} \rangle$. If both insertions are made into the same gluon line, this means inserting the chiral-to-chiral propagator $\la W(5)W(6)\ra \sim \delta(5,6)$ into that gluon line. By performing the chiral superspace integration over point 6, one can show that the five-point
correlator arising in this way precisely cancels against the single-insertion term in this formula.}

The most important property of the new six-point amplitude is that it can 
be written in the factorized form
\bea
\langle Q^{(k)}|{\cal L}|{\cal L} \rangle
= \Theta \times A^{(k-2)}(x,u) \, ,
\label{6pt}
\eea
where $\Theta$ is a particular nilpotent six-point superconformal covariant carrying harmonic U(1) charge 2 at points 1 to 4, and R charge 2 at points 5 and 6 (the chiral field strength $W$ has R charge 1 while the HMs carry harmonic U(1) but no R charge). Since the left-handed $\theta_\alpha$ are the only superspace coordinates with positive R charge $1/2$, we conclude that the expansion of $\Theta$ must start with four such $\theta$s. The rest of the six-point correlator (\ref{6pt}) is given by the function $A^{(k-2)}(x,u)$ which carries U(1) charge $k-2$ at points 1 to 4 and can be expanded in the propagator basis as follows:
\bea
A^{(k-2)}(x,u)=\sum_{m=0}^{k-2}X^{m} Y^{k-m-2} A_m(x) \label{prba}\,  .
\eea
The coefficients $A_m(x)$ are conformally covariant functions of the six space-time points. They have vanishing U(1) charge and are thus harmonic independent.

The structure of the nilpotent covariant $\Theta$ is determined by superconformal symmetry combined with the Grassmann analytic (or $\half$-BPS) nature of the first four points and the chiral nature of the last two points. The explicit form of $\Theta$ (for $\bar\theta^+_r=0$, $r=1,\ldots,4$) has been worked out in Ref. \cite{ESSun}:
\begin{eqnarray}
  \Theta &=& \frac{\prod_{r=1}^4x_{r5}^2x_{r6}^2}{x_{56}^4}\ \frac{x_{12}^2x_{34}^2x_{13}^4x_{24}^4}{R_{{\cal N}=2}} \label{xipsi}\\
  &\times& \left\{[12]^2[34]^2\tau_{14}\tau_{23} +
[14]^2[23]^2\tau_{12}\tau_{34}  +[12][23][34][41]\Bigl[
\tau_{13}\tau_{24}-\tau_{12}\tau_{34}-\tau_{14}\tau_{23}\Bigr]\right\}\,.\nonumber
\end{eqnarray}
Here
\bear
\tau_{rs}&\equiv& 4(\rho_r\rho_s)(\sigma_r\sigma_s)
+\rho_r^2\sigma_s^2 +\rho_s^2\sigma_r^2
\label{deftau}
\ear
and 
\begin{equation}\label{rhosig}
  \rho_r = (\theta_r^+ - \theta_5^i u_{ri}^+) x_{r5}^{-1}\,, \qquad \sigma_r = (\theta_r^+ - \theta_6^i u_{ri}^+) x_{r6}^{-1} \,, \qquad r=1,\ldots,4 
\end{equation}
are Q-supersymmetry invariant combinations of the analytic $\theta_r^+$ with the chiral $\theta_{5,6}^i$ projected with the SU(2) harmonics $(u_r)_i^+$. The harmonic contractions $[12]$, $[34]$, etc. and the polynomial $R_{{\cal N}=2}$
have been defined in eqs. (\ref{harcontra}) and (\ref{RN2}), respectively. 

The aim of our two-loop calculation is to determine the factor $A^{(k-2)}(x,u)$ in the six-point covariant (\ref{6pt}) and then to substitute everything in the insertion formula (\ref{ins}). Since we are only interested in the lowest component of the four-point correlator $\la Q^k \ra$, we can set all the external $\theta$s to zero, $\theta^+_r = \bar\theta^+_r=0$, $r=1,\ldots,4$. In this case $\Theta$ is rather simple:
\begin{equation}\label{otherfr}
  \Theta_{\theta^+=0} = \theta_5^4\theta_6^4\  \frac{x_{12}^2x_{34}^2x_{13}^4x_{24}^4}{x_{56}^4}\ R_{{\cal N}=2}\,.
\end{equation}
Consequently, the six-point correlator becomes
\bea
\label{f6pt}
 \langle Q^{(k)}| {\cal L}|{\cal L}\rangle_{\theta^+=0} = \theta_5^4\theta_6^4\  \frac{x_{12}^2x_{34}^2x_{13}^4x_{24}^4}{x_{56}^4}\ R_{{\cal N}=2}\ A^{(k-2)}(x,u)\, .
\eea
Further, substituting this into the double-insertion formula (\ref{ins}) and performing the trivial chiral integrations over $\theta_{5,6}$, we obtain the two-loop correlator
\bea
 \langle Q^{(k)} \rangle_{\theta^+=0} =
 x_{12}^2x_{34}^2x_{13}^4x_{24}^4~ R_{   {\cal N}=2  }
\int \frac{\rmd^4x_5 \rmd^4x_6}{x_{56}^4}~ A^{(k-2)}(x,u) \, .
\label{famp1}
\eea
Rewriting the amplitude in the form (\ref{pp}), we can read off
the following expression for the conformally invariant coefficient 
functions $\cF^{(k)}$ at two loops:
\bea
\label{final}
\cF_{m,k-m-2,0}^{(k)}(s,t)= x_{12}^2x_{34}^2x_{13}^4x_{24}^4~
\int \frac{\rmd^4x_5 \rmd^4x_6}{x_{56}^4}~ A_m(x) \, .
\eea
                                                    
Now, the practical question is how to compute $A^{(k-2)}(x,u)$ (and hence $A_m(x)$ and $\cF(s,t)$) from the corresponding set of two-loop Feynman diagrams. It turns out that instead of setting $\theta^+=0$, as required in the final expression (\ref{famp1}), it is much more convenient to do the computations with $\theta_{5,6}=0$. The knowledge of the structure of $\Theta$ (\ref{xipsi}) allows us to easily switch from one of these superconformal frames to the other.\footnote{By ``superconformal frames" we mean that $\Theta$ can be cast into one of these forms by means of a finite superconformal transformations.} In the new frame $\Theta$ becomes even simpler,
\bea
\Theta_{\theta_{5,6}=0}=\prod_{r=1}^4
 (\theta^+_r)^2\, ,
\eea
so that eq. (\ref{f6pt}) is replaced by 
\begin{equation}\label{chk}
 \langle Q^{(k)}| {\cal L}|{\cal L}\rangle_{\theta_{5,6}=0} = \prod_{r=1}^4
 (\theta^+_r)^2\ A^{(k-2)}(x,u) \, .
\end{equation}   
Then it is clear that in working out the expressions for the various Feynman graphs we can concentrate only on the terms with the maximal number of external $\theta$s. 
In particular, at order $g^4$ this choice removes all graphs which contain non-Abelian interaction vertices. For example, the Y-shaped gluon subgraph in Fig. 5a  vanishes because it has two chiral ends at the insertion points 5 and 6 and one analytic end (the gluon without insertion); after setting $\theta_{5,6}=0$ we are left with too few left-handed $\theta$s at the analytic gluon end to supply the required R charge 2. Similarly, the TTT block in Fig. 5a has three chiral ends (in fact, only two, points 6 and 7 should be identified) and two analytic ends; once again, the analytic $\theta$s cannot provide the required R charge 3. 

\vskip 15pt
\noindent
\begin{minipage}{\textwidth}
\begin{center}
\includegraphics[width=0.70\textwidth]{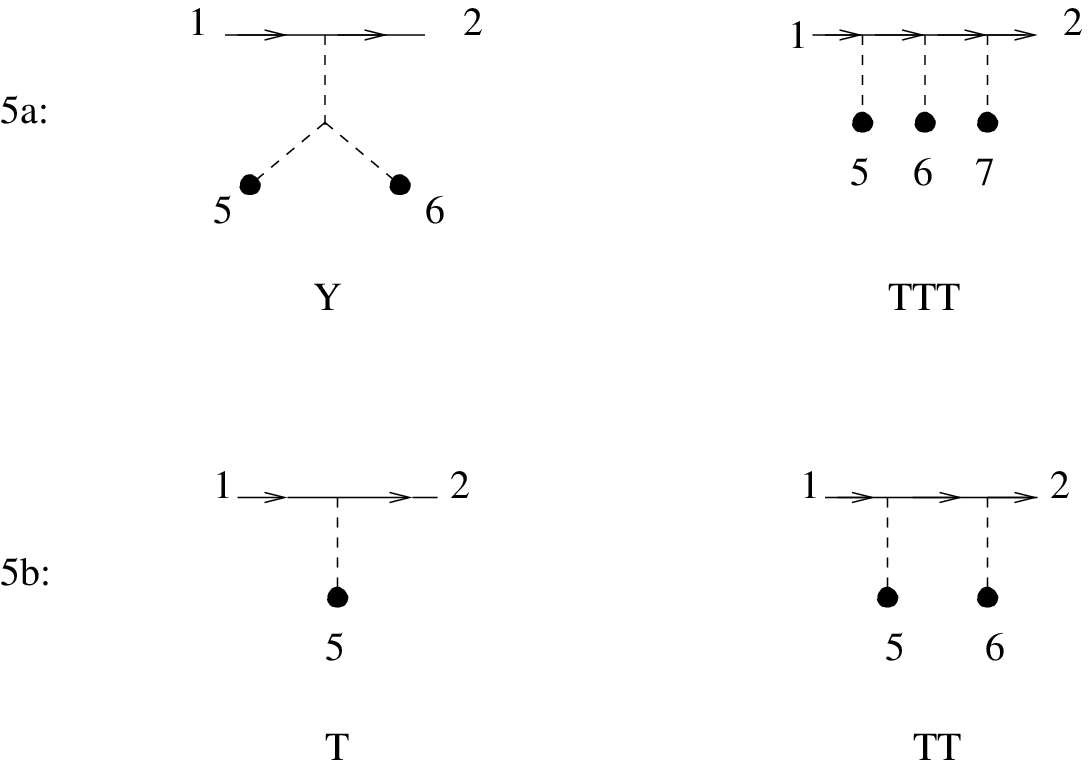}
\end{center}
\begin{center}
Figure 5. Building blocks of the Feynman graphs. 
\end{center}
\end{minipage}
\vskip 20pt

As a result of all these simplifications our task is reduced to listing all {\it tree level} Feynman graphs made out of the two building blocks T and TT in Fig. 5b.  Although these blocks contain interaction vertices and hence integrals, the latter are easily computed using the harmonic superspace Feynman rules and they produce very simple {\it rational space-time functions} \cite{Howe:2000hz,ESSun}: 
\begin{equation}\label{T}
  T_{125} \equiv \langle \widetilde q^+(1) W(5) q^+(2) \rangle = \frac{2gf_{abc}}{(2\pi)^4\;  x_{12}^2}\; \Big[[21^-]\rho_1^2+[12^-]\rho_2^2 -2(\rho_1\rho_2)\Big]\, ,
\end{equation}
\begin{equation}\label{TT}
   TT_{1562} \equiv \langle \widetilde q^+(1) W(5) W(6)  q^+(2) \rangle \, = \, -\frac{4g^2 f_{abc}f_{cde}}{(2\pi)^6\; x_{12}^2} \, [1^-2^-]\rho_1^2\, \sigma_2^2 \, ,
\end{equation}
where $\rho,\sigma$ have been defined in (\ref{rhosig}) and $[12^-]= (u_1)_i^+ \epsilon^{ij} (u_2)_j^-$ (cf. (\ref{harcontra})).

Notice the characteristic presence of negative-charged harmonics in both expressions (\ref{T}) and (\ref{TT}). This has to do with the important issue of harmonic analyticity \cite{Eden:2000qp}. The point is that the free HM satisfies the massless field 
equation
\begin{equation}\label{HMfe}
  D^{++} q^+(x,\theta^+,\bar\theta^+,u) = 0 \, ,
\end{equation}
where
\begin{equation}\label{D++}
  D^{++} = u^{+i}{\partial\over\partial u^{-i}}
-4i\theta^{+}\sigma^\mu\bar\theta^{+} {\partial\over\partial
x^\mu} 
\end{equation}
is the supersymmetrized harmonic derivative (the raising operator of the group SU(2) realized on the charges $\pm$ of the harmonics). This equation can also be viewed as a Cauchy-Riemann condition on the harmonic coset SU(2)/U(1), hence the name ``harmonic analyticity". Yet another interpretation of eq. (\ref{HMfe}) is that it defines $q^+$ as the highest weight state of an SU(2) irrep of charge 1 (isodoublet). When the HMs interact with the gauge sector, the harmonic derivative in (\ref{HMfe}) is modified by a gauge connection. Still, the gauge invariant composite operators like ${\rm Tr} (q^+)^k$ satisfy the same equation with a flat harmonic derivative. So, such operators correspond to the highest weight state of an SU(2) irrep of charge $k$.   

As the simplest example of harmonic analyticity in action, consider the HM two-point function (propagator). It has to satisfy the Green's function equation
\begin{equation}\label{HMproeq}
  D^{++}_1\langle \widetilde q^+(1) q^+(2)\rangle = \delta(1,2)\,,
\end{equation}
so here harmonic analyticity holds up to contact terms. 
The solution to this equation is (for simplicity we set $\theta_2=0$)
\begin{equation}\label{HMpro}
   \langle \widetilde q^+(1) q^+(2)\rangle = \frac{[12]}{\left(x_{12} + 4i \frac{[1^-2]}{[12]} \theta^{+}_1\sigma\bar\theta^{+}_1\right)^2} \,.
\end{equation}
This can be checked by expanding the denominator in $\theta^+_1\bar\theta^+_1$, using the relation $\square_1 (1/x^2_{12}) \sim \delta(x_{12})$ and the naive harmonic differentiation rules
\begin{equation}\label{harde}
  D^{++}_1\theta^+_1 = 0, \quad \quad D^{++}_1[1^-2] = [12], \quad D^{++}_1 [12] = 0 \,,
\end{equation}
as well as the rule
\begin{equation}\label{singder}
  D^{++}_1 \frac{1}{[12]} = \delta(u_1,u_2)
\end{equation}
which follows from the relation $\pa/\pa \bar z (1/z) \sim \delta(z)$. 
In the context of our four-point investigations we always set all 
$\bar\theta=0$. Therefore here we never encounter harmonic singularities 
of the type (\ref{singder}) and can safely apply the naive rules 
(\ref{harde}). Then harmonic analyticity simply means {\it polynomial 
dependence in $u^+$ and no dependence in $u^-$}. This is the case of 
the propagator (\ref{HMpro}) which now becomes 
\begin{equation}\label{propth}
\langle \widetilde q^+(1) q^+(2)\rangle_{\bar\theta=0} = 
\frac{[12]}{x^2_{12}} \,.
\end{equation}
The same must hold for any $n$-point correlator of $\half$-BPS operators: 
They must be polynomials in the harmonic variables $u^+$ at each point.

Clearly, the expressions for the building blocks T (\ref{T}) and 
TT (\ref{TT}) are 
not harmonic analytic because of the presence of $1^-$ and $2^-$. 
This, however, 
is not a problem: The various building blocks or even complete Feynman 
graphs are 
not expected to be harmonic analytic, much like they are not conformal 
covariants. 
It is only the sum of all graphs that has these properties 
(see examples in Appendix A). 
In what follows we shall often profit from the expected harmonic analyticity 
of the final result to greatly simplify our graph calculations.

\section{Two-loop four-point amplitudes}

In this section we work in the frame $\theta_{5,6}=0$ where the six-point 
nilpotent 
covariant takes the form
\begin{eqnarray}
 \langle Q^{(k)}| {\cal L}|{\cal L}\rangle_{\theta_{5,6}=0}\ &=&\prod_{r=1}^4
 (\theta^+_r)^2 \label{complamk}
\left[\sum_{m=0}^{k-2}
X^{m} Y^{k-m-2} A_m(x)  \right]\, . \label{Am}
\end{eqnarray}
Here the coefficients $A_m(x)$ are not all independent, as the crossing 
symmetry 
$2\leftrightarrow 3$
of eq. (\ref{pp})  implies that $A_m(x)$ is equal to
$A_{k-m-2}(x)$ with $x_2$ and $x_3$ interchanged.
In particular, for $k$ even, $A_{\frac{k-2}{2}}$ is invariant
under this crossing transformation. By using eq. (\ref{final})
one can see that these facts are in accord with
the corresponding transformation property of $\cF_{m,k-m-2,0}^{(k)}(s,t)$.
Therefore, we need compute only $(\frac{k-2}{2}+1)$ coefficient functions 
if $k$ is even and $\frac{k-1}{2}$ if $k$ is odd.

At two loops the correlator (\ref{Am}) is given by the sum of a number 
of Feynman graphs 
containing four interaction vertices of HM with gluons and two ${\cal L}$ 
insertions.
Let us discuss the general structure of these diagrams. According to the 
rules discussed
in section 4, they are made out of the building blocks T and TT.
The same argument based on $\theta$ counting allows us to discard other sets
of vanishing diagrams. Take, for instance, the graphs involving a 
HM loop with 
the insertion of two gluon lines connecting the HM propagators. Such a 
loop produces the expression $(TT)^2 \sim \rho^4 \sigma^4 = 0$ 
(see (\ref{TT})). 
For the same reason any graph containing a corner with all the incoming or 
outgoing HM 
lines being free vanishes, since this implies concentrating too many 
$\theta$s at some 
other corner. Finally, disconnected graphs containing subgraphs in the 
form of the one- 
or two-loop corrections to the two-point function of protected 1/2 BPS 
operators vanish as well. 

With these rules in mind and observing that at order $g^4$ the 
maximal number of 
HM lines involved in interactions with gluons is four, we can now draw 
all possible 
configurations. The planar interaction topologies {\it relevant} to the 
calculation of 
the functions $A_m(x)$ are the same as the ones shown in Fig. 12 for 
the $k=4$ case (thin lines).\footnote{Beyond those of Fig. 12
there are also other interacting topologies which, however, play only an 
auxiliary r\^ole as it will become clear from our procedure of 
identifying the diagrammatic contributions to $A_m(x)$.} 
The complete graphs are then obtained by 
combining an interaction 
topology from Fig. 12 (thin lines) with the matching free HM frame 
from Fig. 6a. 
Each resulting topology is labeled by a pair of integers $(p,q)$ 
($0\leq p\leq 2m$, $0\leq 
q\leq 2(k-m-2)$) which fixes the particular structure of the free HM lines.
This procedure automatically takes into account the possible crossing symmetry 
transformations of the end points.  

\vskip 15pt
\noindent
\begin{minipage}{\textwidth}
\begin{center}
\includegraphics[width=0.6\textwidth]{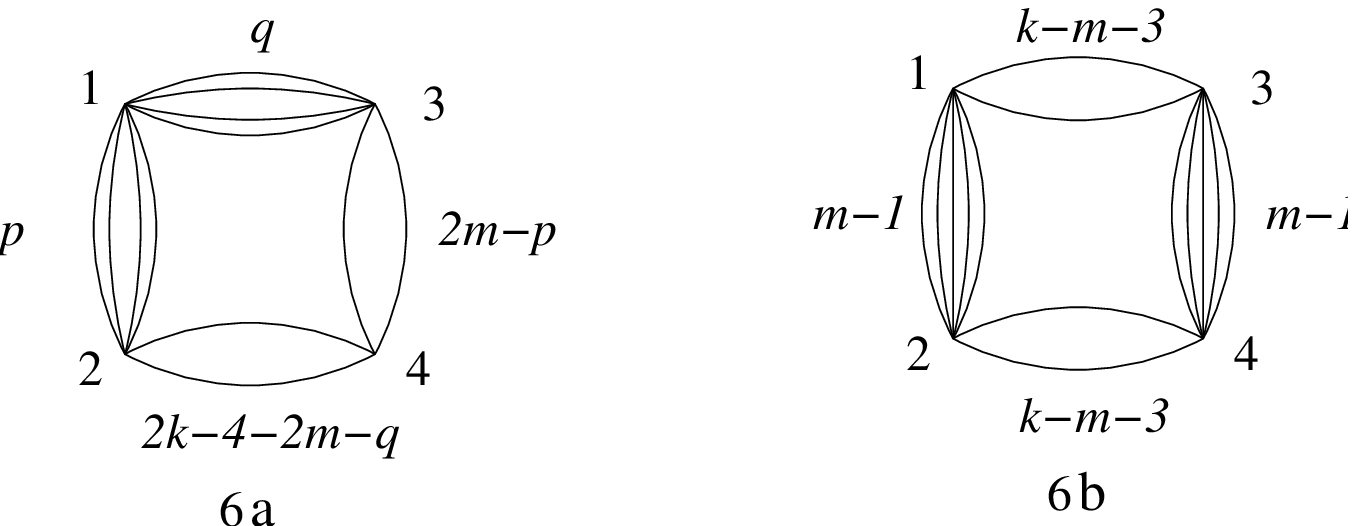}
\end{center}
\begin{center}
Figure 6. a) The free HM frame $(p,q)$ associated with the function $A_m(x)$.
b) The HM frame common to all graphs contributing to $A_m(x)$. 
\end{center}
\end{minipage}
\vskip 20pt

\noindent

Evaluating the explicit expressions for each topology amounts to multiplying 
the building blocks (\ref{T}) and (\ref{TT}) and using a simple Fierz 
identity for the 
two-component spinors $\rho$ and $\sigma$. In this way the contributions of 
the individual 
graphs are always expressed in terms of the variables $\rho_r^2$, 
$\sigma_r^2$ and $\tau_{rs}$. 

Since the building blocks contain negatively charged harmonic variables,
each individual graph involves some non-analytic harmonic
dependence. The final sum (\ref{Am}) must however be harmonic analytic
and moreover, the coefficients $A_m$ are harmonic independent.
Thus, all the terms in the graphs with non-analytic dependence are in fact
spurious and should cancel out. Due to the huge number of diagrams the
evaluation of each diagram and the proof of the actual cancellation
of non-analytic terms can be rather intricate (see Appendix A for a sample 
calculation). 

However, the knowledge that the final result has to be analytic
allows us to skip the actual computation of all the diagrams 
and directly identify a minimal set of graphs relevant to restoring the 
correct harmonic analytic structure of the final sum. This can be achieved
by a suitable procedure of identifying the harmonics. In Appendix B we describe 
this procedure in detail when applied to operators of weight $k=3,4$. 
The corresponding minimal sets of relevant graphs for the $k=3$ case are given
in Fig. 11, whereas for $k=4$ they are listed in Figs. 13 and 14.

For generic $k$, the procedure of identifying the harmonics goes along the
same lines as in Appendix B. In this way, we are led to select the minimal set of relevant 
diagrams which contribute to the function $A_m$. They are drawn in Fig. 7 
where the labels $(p,q)$ are explicitly indicated for each graph.

\vskip 20pt
\noindent
\begin{minipage}{\textwidth}
\begin{center}
\includegraphics[width=0.7\textwidth]{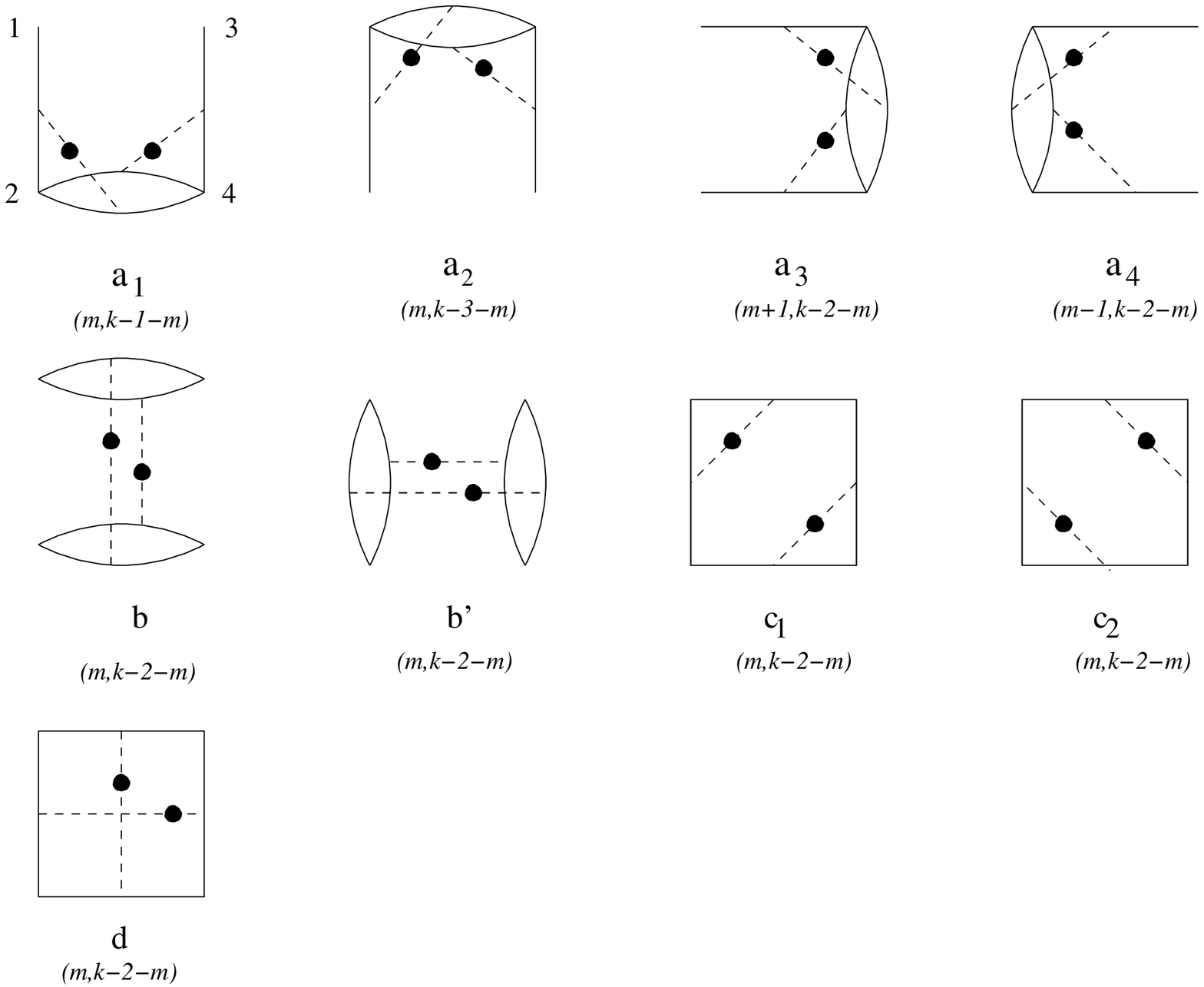}
\end{center}
\begin{center}
Figure 7. Relevant topologies of the interacting subgraphs of the planar 
graphs contributing
to the function $A_m$. Every subgraph has to be multiplied by the free 
contribution from Fig. 6a.
\end{center}
\end{minipage}
\vskip 20pt

\noindent

We note that the topologies of interaction lines appearing in the diagrams 
of Fig. 7 are the same as the ones for $k=3,4$ cases (see Figures 11--14
in Appendix B). 
Let us explain why this happens by first looking at the case 
$1\leq m\leq k-3$. 
We observe that the diagrams in Fig. 7 all contain the common free HM frame 
depicted in Fig. 6b. This frame provides the harmonic factor
\bea
\label{com}
X^{m-1} Y^{k-m-3} \rightarrow 
\Big([12][43]\Big)^{m-1}\Big([13][42]\Big)^{k-m-3} 
\eea
which can be pulled out from all the graphs. It is easy to see
that what remains forms the set of graphs contributing to the 
function $A_1(x)$ for the amplitude of weight 4 operators (see Fig. 14). 
Therefore, we can exploit the results for $k=4$ to select  
the minimal set of interacting subdiagrams which survive harmonic 
identification.  
The particular cases $m=0$ and its crossing symmetry partner $m= k-2$ can be 
treated
in the same way. For $m=0$ the graphs a$_3$ and a$_4$ in Fig. 7 are absent and 
pulling out a common factor $Y^{k-4}$ we are led to the set of graphs for the
function $A_0$ of the $k=4$ case (see Fig. 13). 

The important conclusion we reach at this point is that {\it the two-loop
calculation for arbitrary
$k$ is reduced to that for $k=4$}. The reason for this is quite clear:
Since at two loops at most four HM lines are involved in the interactions, 
for a sufficiently large $k$ the majority of the HM lines play the r\^ole of 
``spectators".

Let us explain why the diagrams in Fig. 7 with the 
labels $(p,q)$ indicated do indeed provide the contributions to $A_m$.  
Recall that the graphs which contribute 
to $A_m$ are the ones from which we can eventually extract the propagator 
factor $X^{m} Y^{k-m-2}$. We have already pulled out a common factor 
$X^{m-1} Y^{k-m-3}$.
Leaving aside for the moment the graphs of the type $7$a we see that  
all the other diagrams contain an additional factor 
$XY\sim [12][43][13][42]$. 
Pulling it out we restore the required overall factor $X^{m}Y^{k-m-2}$. 
What remains from the graph is a chargeless combination of harmonics which 
may contain constant harmonic independent terms, 
but also non-analytic terms. However, once we have pulled out 
the required overall factor, the rest has to be harmonic independent
and can be computed by identifying all the harmonics as explained in 
Appendix B.  

We still have to take care of the graphs which contain interaction 
subgraphs of type $7$a. 
For instance, for the interaction topology $7$a$_1$ besides the common 
frame (\ref{com}) 
we can also pull out an extra factor $[12][43] \ \sim X$. This leaves 
behind the free 
factor $[13]^2$ which is different from the expected $[13][42]\ \sim \ Y$. 
The missing 
factor $Y$ is restored only after summing up $7$a$_1$ with all of its 
analyticity partners. 
The partners of the graph $7$a$_1$ needed to restore analyticity,
e.g., at points 1 and 3 are shown in Fig. 8. In Appendix A we analyze 
in detail how 
analyticity is achieved in this case and how such a long calculation can be 
efficiently bypassed by the procedure of identifying the harmonics.

\vskip 20pt
\noindent
\begin{minipage}{\textwidth}
\begin{center}
\includegraphics[width=0.8\textwidth]{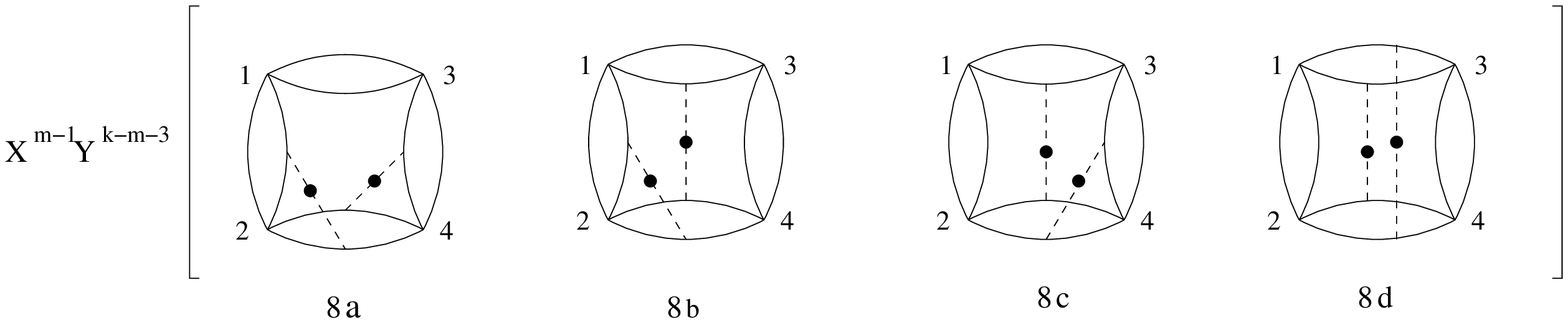}
\end{center}
\begin{center}
Figure 8. The graph $8$a (topology $7$a$_1$)
and its partners restoring analyticity at points 1 and 3 (all the 
graphs have the same bottom part). The common free HM
frame has been pulled out as the prefactor $X^{m-1}Y^{k-m-3}$. 
\end{center}
\end{minipage}

\vskip 20pt \noindent

The reader may wonder why the graph 8d, which has the topology 7b, 
has been included 
in the set 8 where it clearly plays the auxiliary role of an analyticity 
partner of the graph 8a. In fact, this graph plays a dual role: 
It becomes the principal graph contributing to the 
coefficient function $A_{m+1}(x)$. Indeed, pulling out the common factor 
$X^{m-1} Y^{k-m-3}$ 
and the extra factor $X^2$ contained in the the graph 8d, we find 
\bea \label{a8d}
\Bigl[\tau_{13}-[13] [1^-3^-] (\rho^2_1\sigma^2_3 +
\rho^2_3\sigma^2_1)\Bigr]\ \Bigl[\tau_{24} -[24][2^-4^-](\rho^2_4\sigma^2_2 +
\rho^2_2\sigma^2_4)\Bigr] \, .
\eea
This expression is chargeless, so in it we can identify all the harmonics, 
after which we are 
left with the analytic term $\tau_{13}\tau_{24}$. We can say that the 
analytic part of 
eq. (\ref{a8d}) contributes to the function $A_{m+1}$, while the non-analytic 
part combines 
with the other graphs in Fig. 8 into a harmonic analytic expression which 
gives a contribution to $A_m$.  

Having identified the relevant graphs, we now proceed to computing the 
six-point correlator. The Grassmann and space-time structures of the graphs 
in Fig. 7 are
\bea
\nonumber
G_{a_1}&=&\Big(T_{125}T_{346}+(5\leftrightarrow 6)\Big)T_{245}T_{246}=
\frac{\t_{24}(\r_1^2\s_3^2+\r_3^2\s_1^2) }{x_{24}^4x_{12}^2 x_{34}^2} \, , \\
\nonumber
G_{a_2}&=&\Big(T_{125}T_{346}+(5\leftrightarrow 6)\Big)T_{135}T_{136}=
\frac{ \t_{13}(\r_2^2\s_4^2+\r_4^2\s_2^2)}{x_{13}^4x_{12}^2 x_{34}^2}\, , \\
\nonumber
G_{a_3}&=&\Big(T_{135}T_{246}+(5\leftrightarrow 6)\Big)T_{345}T_{346}=
\frac{ \t_{34}(\r_1^2\s_2^2+\r_2^2\s_1^2)}{x_{13}^2x_{24}^2 x_{34}^4}
\, , \\
\nonumber
G_{a_4}&=&\Big(T_{135}T_{246}+(5\leftrightarrow 6)\Big)T_{125}T_{126}=
\frac{ \t_{12}(\r_3^2\s_4^2+\r_4^2\s_3^2)}{x_{12}^4x_{13}^2x_{24}^2 }
\, , \\
\nonumber
G_{b}&=&
 T_{135}T_{136}T_{245}T_{246}+(5\leftrightarrow 6)=
2\frac{\t_{13}\t_{24} }{x_{13}^4 x_{24}^4} \, ,\\
\nonumber
G_{b'}&=&T_{125}T_{126}T_{345}T_{346}+(5\leftrightarrow 6)=
2\frac{ \t_{12}\t_{34} }{x_{12}^4 x_{34}^4} \, ,\\
\nonumber
G_{c_1}&=&
T_{125}T_{135}T_{246}T_{346}+(5\leftrightarrow 6) =
\frac{\t_{23}(\r_1^2\s_4^2+\r_4^2\s_1^2)}{x_{13}^2x_{24}^2x_{12}^2 x_{34}^2} \\
\nonumber
G_{c_2}&=&
T_{125}T_{245}T_{136}T_{346}+(5\leftrightarrow 6) =
\frac{\t_{14}(\r_2^2\s_3^2+\r_3^2\s_2^2)}{x_{13}^2x_{24}^2x_{12}^2 x_{34}^2} \, ,
\eea
and
\bea
\non
G_d&=&
T_{125}T_{345}T_{136}T_{246}+(5\leftrightarrow 6)\\
\nonumber
&=&
\frac{1}{x_{12}^2x_{34}^2x_{13}^2x_{24}^2}
\Big[
\t_{14}\t_{23}-\t_{13}\t_{24}-\t_{12}\t_{34}
-\t_{23}(\r_1^2\s_4^2+\r_4^2\s_1^2)-\t_{14}(\r_3^2\s_2^2+\r_2^2\s_3^2)\\
\nonumber
&+&\t_{13}(\r_4^2\s_2^2+\r_2^2\s_4^2)+\t_{24}(\r_1^2\s_3^2+\r_3^2\s_1^2)+
\t_{12}(\r_4^2\s_3^2+\r_3^2\s_4^2)+\t_{34}(\r_1^2\s_2^2+\r_2^2\s_1^2)
\Big] \, .
\eea
Collecting everything we arrive at the following complete harmonic 
analytic expression for the six-point correlator: 
\bea 
\label{6ptcor}
\langle Q^{(k)}|{\cal L}|{\cal
L}\rangle
&\sim &\frac{1}{x_{12}^2x_{34}^2x_{13}^2x_{24}^2}\sum_{m=0}^{k-2} 
X^{m}Y^{k-m-2}\\
\nonumber
&\times&
\Big\{ C^{d}_m \t_{14}\t_{23}+
\Big(2sC_m^{b} -C_m^d\Big)\t_{13}\t_{24}+
\Big(\frac{2}{s}C_m^{b'}-C_m^d\Big)\t_{12}\t_{34}  \\
\nonumber
&+&(C_m^{c_1}-C_m^d)\t_{23}(\r_1^2\s_4^2+\r_4^2\s_1^2)
+(C_m^{c_2}-C_m^d)\t_{14}(\r_3^2\s_2^2+\r_2^2\s_3^2)\\
\nonumber
&+&(C_m^d-C_m^{a_2})\t_{13}(\r_4^2\s_2^2+\r_2^2\s_4^2)
+(C_m^d-C_m^{a_1})\t_{24}(\r_1^2\s_3^2+\r_3^2\s_1^2)\\
\nonumber
&+&(C_m^d-C_m^{a_4}) \t_{12}(\r_4^2\s_3^2+\r_3^2\s_4^2)
+(C_m^d-C_m^{a_3}) \t_{34}(\r_1^2\s_2^2+\r_2^2\s_1^2)
 \Big\} \, .
\eea
where $C_m$ denote the combinatorial factors associated to the various graphs. 

The expression (\ref{6ptcor}) has been derived for $1\leq m\leq k-3$.
As was already mentioned above,
the cases $m=0$ (and its crossing partner $m=k-2$) are a bit special, 
as for $m=0$ the graphs ${\rm a}_3$ 
and ${\rm a}_4$ are absent, while for $m=k-2$ we do not have ${\rm a}_1$ 
and ${\rm a}_2$. 
Otherwise, 
the derivation of the relevant contributions for these cases follows
the same steps as above. Thus, the formula (\ref{6ptcor})
remains valid for $m=0$ and for $m=k-2$ as well, provided we impose 
the following formal requirement on the combinatorial coefficients 
$$
C^{a_3}_0=C^{a_4}_0=C^{a_1}_{k-2}=C^{a_2}_{k-2}=0 \, .
$$

Before we proceed, let us briefly discuss the general symmetry
properties of the combinatorial factors. We note that for
the graphs ${\rm a}_1$ and ${\rm a}_2$ the range of indices is $0\leq m\leq
k-3$, while for ${\rm a}_3$ and ${\rm a}_4$ it is $1\leq m\leq k-2$. For a
generic $m$ the combinatorial coefficients $C_m^{a_1}$ and
$C_m^{a_2}$ corresponding to the graphs ${\rm a}_1$ and ${\rm a}_2$ are
equal, $C_m^{a_1}=C_m^{a_2}\equiv C_m^a$  (though their analytic
structures are different). Analogously,
$C_m^{a_3}=C_m^{a_4}\equiv C_m^{a'}$ and
$C_m^{c_1}=C_m^{c_2}=C_m^c$. We also have the following
additional relations 
\bea C_{m}^a=C_{k-m-2}^{a'}\, ,
~~~~C_{m}^{b}=C_{k-m-2}^{b'}\, , \eea 
which in the special case $m=m_{-}$ reduce to \bea \label{special} C_{m_-}^a=
C_{m_-}^{a'}\, , ~~~~C_{m_-}^{b}=C_{m_-}^{b'}\, . \eea


Finally, to recast the six-point amplitude (\ref{6ptcor}) in the form 
(\ref{Am}) we can use the relations
\bear \rho_r^2 &=& {\theta_r^{+2}\over x_{r5}^2}, \quad \sigma_r^2
= {\theta_r^{+2}\over x_{r6}^2}, \quad \tau_{rs} =
\theta_r^{+2}\theta_s^{+2}{x_{rs}^2x_{56}^2 \over
x_{r5}^2x_{r6}^2x_{s5}^2x_{s6}^2} \label{rhotauexplicit} \ear 
valid for $\theta_{5,6}=0$. The very last step is to perform the 
space-time integration over the insertion points 5 and 6 in eq. (\ref{final}). 
This introduces the functions $\Phi^{(1)}$ and $\Phi^{(2)}$ defined
in (\ref{h1}) and (\ref{h2}) and their point permutations. 
Note that  $\Phi^{(1)}$ is totally 
symmetric under permutations of the external points.

With these definitions at hand, from eqs. (\ref{Am}) and (\ref{final})  
we can identify
\bea
\label{G}
&&\cF_{m,k-m-2,0}^{(k)}(s,t)=\varkappa\Big[ \frac{1}{4}
\Big(C_m^dt+(2C_m^{b} -C_m^d)s+(2C_m^{b'}-C_m^d)
\Big) [\Phi^{(1)}(s,t)]^2\\
\nonumber &&~~~+(C_m^d-C_m^{a'})\frac{1}{s}\Phi^{(2)}(t/s,1/s)
+(C_m^d-C_m^{a})\Phi^{(2)}(s,t)
+(C_m^{c}-C_m^d)\frac{1}{t}\Phi^{(2)}(s/t,1/t) \Big]\, . \eea
Here $\varkappa$ is an overall normalization coefficient independent of 
$m$, which is 
related to the normalization of the $\half$-BPS operators.

The final result (\ref{G}) essentially depends on the values of
the combinatorial coefficients $C_m$. The particular crossing symmetry 
properties 
(\ref{pmcross}) of the functions $\cF_{\pm}$ provide some additional 
relations among the 
corresponding combinatorial coefficients. Indeed,
for $m=m_+$ we find
\bea
C_{m_+}^c=2C_{m_+}^{b'}=2C_{m_+}^{d}\, ~~~~~
\eea
but there are no new restrictions on $C_{m_-}$.

We remark that the case $k=2$ is special as there are no  graphs of type 
$6$a and the higher 
crossing symmetry requires $2C^{b}_{0}=C_0^c=2C_0^d$.

The general expressions for the combinatorial coefficients $C_m$
valid for finite $N$ are 
\bea
\nonumber
C_m^a&=&
\frac{f_{qpe}f_{rte}f_{cse'}f_{nle'}}{2(m!)^2(k-1-m)!(k-3-m)!}
(a_1...a_ma_{m+1}...a_{k-1}p)
(a_1...a_mb_{m+1}...b_{k-3}qrn)\\
\nonumber
&\times&(c_1...c_ma_{m+1}...a_{k-1}c)
(c_1...c_mb_{m+1}...b_{k-3}tsl) \, ,  \\
\nonumber
2C_m^b&=&\frac{f_{qpe}f_{rte}f_{cse'}f_{nle'}}{2[(k-m-2)!m!]^2}
(a_1...a_ma_{m+1}...a_{k-2}ps)
(a_1...a_mb_{m+1}...b_{k-2}rn)\\
\nonumber
&\times&(c_1...c_ma_{m+1}...a_{k-2}qc)
(c_1...c_mb_{m+1}...b_{k-2}tl) \, ,  \\
\nonumber
C_m^c&=&\frac{f_{qpe}f_{rte}f_{cse'}f_{nle'}}{2[(k-m-2)!m!]^2}
(a_1...a_ma_{m+1}...a_{k-2}pt)
(a_1...a_mb_{m+1}...b_{k-2}qn)\\
\nonumber
&\times&(c_1...c_ma_{m+1}...a_{k-2}rc)
(c_1...c_mb_{m+1}...b_{k-2}sl) \, ,  \\
\nonumber
C_m^d&=&\frac{f_{qpe}f_{rte}f_{cse'}f_{nle'}}{2[(k-m-2)!m!]^2}
(a_1...a_ma_{m+1}...a_{k-2}cp)
(a_1...a_mb_{m+1}...b_{k-2}lq)\\
\nonumber
&\times&(c_1...c_ma_{m+1}...a_{k-2}sr)
(c_1...c_mb_{m+1}...b_{k-2}nt) \, ,
\eea
where
\bea
\nonumber
(a_1\ldots a_k)\equiv \mbox{Tr}\Big(t_{(a_1}\ldots   t_{a_k)} \Big)
\eea
is the symmetrized (without $1/k!$) trace of $k$ generators of the color group.
In these formulae the combinatorial factors are necessary to prevent the
overcounting of the HM lines induced by  symmetrization.

To compute these coefficients in the large $N$ limit we use the
conventions  $[t_p,t_q]=if_{pqe}t_{e}$  and
$f_{abe}f_{abe'}=2N\d_{ee'}$ as well as the following fusion and
splitting rules \bea \nonumber
(a_1\ldots a_lA)(a_1\ldots a_l B) &\approx & (l+1)^2l!N^{l-1}\mbox{Tr}(AB) \, 
,\\
\nonumber
\mbox{Tr}[a_l\ldots a_1a_1\ldots a_lA] &\approx & N^l \mbox{Tr}A \, .
\eea
In the large
$N$ limit we find
\bea
C_m^a=2C_m^b=C_m^d=k^4 N^{2k} \, , ~~~~~~~C_m^c=2k^4N^{2k} \, .
\eea 
In this formula the range of the index $m$ is $0\leq m\leq k-3$ for $C_m^a$, 
$0\leq m\leq k-2$ for $C_m^d$, $1\leq m\leq k-2$ for $C_m^b$ and 
$0\leq m\leq k-2$ 
for $C_m^c$. The case of $C_0^b$ deserves some special attention. 
Looking at the graph 6b 
for $m=0$, we realize that the gluon lines can ``slide" along the HM 
loops, thus giving rise 
to two equivalent planar diagrams. Hence, the factor $C_0^b$ is twice 
as big as the 
other $C_m^b$ with $m\neq 0$:
\bea 
C_0^b=k^4N^{2k}\, \, .
\eea 
So, the combinatorial coefficients have the remarkable property that they do
not depend on $m$, except for the single ``anomalous'' coefficient
$C_m^b$ which changes its value only for $m=0$.

Substituting these combinatorial coefficients into eq. (\ref{G}), 
choosing canonical  normalization for the $\half$-BPS operators and 
recalling the definition  (\ref{Newparam}), we obtain our main result 
(\ref{minpl}).

\section{Two-loop four-point amplitudes from the OPE}

Here we show that the four-point {\it two-loop amplitude} for $\half$-BPS operators of weight 3 can be completely reconstructed just on the basis of the expected operator product expansion structure and the knowledge of the {\it one-loop anomalous dimensions} for certain operators of twist 2 and 4.  Analogously, we obtain some restrictions on the structure of the amplitude for weight 4 operators. The discussion in this section provides an independent check on our diagrammatic computation.

\subsection{Weight 3}
The diagrammatic treatment clearly shows that at two loops the four-point amplitude is made out of two basic conformal integrals, (\ref{h1}) and (\ref{h2}). Therefore, the most general Ansatz compatible with the symmetry requirement
${\cal F}(s,t)=1/t~{\cal F}(s/t,1/t)$ is
\begin{eqnarray}\label{2loop3}
  \cF(s,t)&=&\frac{\lambda^2}{N^2}\cdot\Big[\frac{1}{4}\Big(\varrho s+\eta (t+1)\Big)[\Phi^{(1)}(s,t)]^2 \\
\nonumber
&+&
 \frac{\mu}{s}\Phi^{(2)}(t/s,1/s)+ \nu\Big(\Phi^{(2)}(s,t)
+ \frac{1}{t}\Phi^{(2)}(s/t,1/t)\Big)
 \Big] \, ,
\end{eqnarray}
i.e. each function depends on four unknown coefficients $\varrho,\eta, 
\mu$ and $\nu$.

Already at the level of the free field theory one can see that the OPE ${\cal O}^{(k)}{\cal O}^{(k)}$ of two $\half$-BPS operators exhibits a heredity property: If some superconformal primary operator emerges in the spectrum of this OPE then it must also appear in the spectrum of ${\cal O}^{(n)}{\cal O}^{(n)}$ with $n > k$. Therefore, knowing the spectrum of the operator product ${\cal O}^{(2)}{\cal O}^{(2)}$ (see Refs. \cite{AFP,AEPS}), we can use it to obtain restrictions on the four-point amplitude underlying the OPE of the $\half$-BPS operators with higher weights. From the technical point of view, we need to build the conformal partial wave amplitude (CPWA) expansion of the conformal integrals entering eq. (\ref{2loop3}) and then to match the corresponding short-distance expansion of the four-point amplitude with perturbative CPWA contributions of operators with known anomalous dimensions. In this way one can fix the undetermined numerical coefficients in eq. (\ref{2loop3}).

The CPWA expansions of the conformal integrals in  eq. (\ref{2loop3}), corresponding to taking a short-distance limit $x_1\to x_2$, $x_3\to x_4$, have already been obtained in Ref. \cite{AEPS} (cf. Section 3 there). They are formulated in terms of the conformal cross-ratios $v=s/t$ and $Y=(t-1)/t$. Under the assumption that the conformal dimension  $\Delta$ of an operator is of the form $\Delta=\Delta_0+\gamma$, where $\Delta_0$ and $\gamma$ are the canonical and the anomalous parts, respectively, the non-analytic term $v^{\Delta/2}$ entering the CPWA of this operator gives rise to perturbative logarithms
$$
v^{\Delta/2}=1+1/2~\gamma\ln v+1/8~\gamma^2\ln^2 v+\ldots
$$
Therefore, the information about the one-loop (order $\lambda$) anomalous dimensions is reflected in the coefficients of the $\ln ^2 v$ terms in the perturbative CPWA expansions and this is how we can use it to derive the form of the two-loop (order $\lambda^2$) amplitude. The explicit calculations are not particularly instructive and therefore in the following we mainly restrict ourselves to the formulation of the results obtained.

\vskip 0.3cm
{\it Operators of twist 2.}
\vskip 0.3cm

The R-symmetry singlet of the OPE ${\cal O}^{(2)}{\cal O}^{(2)}$ contains an infinite tower of twist 2 operators of increasing spin, the lowest member being the Konishi scalar of canonical dimension 2 and of one-loop anomalous dimension $3\lambda$. The requirement of reproducing this field in the OPE derived from eq. (\ref{2loop3}) fixes two of the four unknown coefficients, namely,
\bea
\label{K3}
\eta=\nu=\frac{3^2}{4} \, .
\eea
Given relation (\ref{K3}), the whole tower of twist 2 operators is then correctly reproduced by the four-point amplitude. However, the operators of twist 2 produce no restrictions on the coefficients $\varrho$ and $\nu$.

\vskip 0.3cm
{\it Operators of twist 4 in the singlet channel.}
\vskip 0.3cm

To get further restrictions on the four-point amplitude we consider the superconformal primary operators of twist 4 in the singlet R-symmetry channel. The simplest example are the four quadrilinear dimension 4 operators $\Sigma_{1,2}, \Sigma_{\pm}$ studied in Ref. \cite{APPSS} (see also \cite{BKS}). 
These operators diagonalize the matrix of the dilatation operator at one loop with the following result for the large $N$ one-loop anomalous dimensions\footnote{At fixed $\lambda$ the anomalous dimension $\gamma_1$ vanishes only when $N=\infty$ \cite{APPSS}.}
\bea
\label{dg}
\gamma_1=0\, , ~~~
\gamma_2=6\lambda \, , ~~~~
\gamma_{\Sigma_{\pm}}=\frac{1}{4}(13\pm \sqrt{41})\lambda\, . ~~~~
\eea
By using the explicit form of the (canonically normalized) operators $\Sigma_{1,2}$, $\Sigma_{\pm}$ established in Ref. \cite{APPSS} we then compute their free three-point functions with the canonically normalized $\half$-BPS operator ${\cal O}^{(3)}$ in the large $N$ limit  and find
\bea
\label{nc}
 \sqrt{A_{\pm}}=\langle {\cal O}^{(3)}{\cal O}^{(3)}\Sigma_{\pm} \rangle
=\frac{1}{N}\left(\frac{3}{5}\mp\frac{18}{5\sqrt{41}}\right)^{\frac{1}{2}}\, ,
\eea
while $\langle {\cal O}^{(3)}{\cal O}^{(3)}\Sigma_{1,2} \rangle\sim O(1/N^2)$. Such a suppressed behavior of the three-point amplitudes involving $\Sigma_{1,2}$ is naturally explained by the fact that 
these fields are double-trace operators. 
Thus, among the four operators only $\Sigma_{\pm}$ participate in the large $N$ operator product expansion of the four-point amplitude. The normalizations of their CPWAs are given by the constants $A_{\pm}$.

Combining the free, the one-loop amplitudes \cite{AS} and the two-loop Ansatz (\ref{2loop3}), we have worked out the corresponding CPWA expansion and have found the following system of equations\footnote{These equations are obtained after careful disentangling of the contribution of the superconformal descendants of the twist 2 fields.}  encoding the large $N$ contribution of the fields $\Sigma_{\pm}$:
\bea
&& A_{+}+A_{-}=\frac{6}{5N^2} \, , \\
&& A_{+}\gamma_{\Sigma_+}+A_{-}\gamma_{\Sigma_-}=\frac{21}{10}\frac{\lambda}{N^2}\, ,\\
&& A_{+}\gamma_{\Sigma_+}^2+A_{-}\gamma_{\Sigma_-}^2=\Big(\frac{81}{20}+\frac{8}{3}\varrho+\frac{4}{3}\mu\Big)\frac{\lambda^2}{N^2}\, .
\eea
It is easy to check that the first two equations are indeed satisfied with the values of the $A$'s and $\gamma$'s found, while the third equation yields
\bea
\varrho=-\frac{\mu}{2}\, .
\eea
At this stage only one coefficient, e.g. $\mu$, remains undetermined.

\vskip 0.3cm
{\it Operators of twist 4 in the irrep $[0,2,0]$.}
\vskip 0.3cm
The OPE ${\cal O}^{(2)}{\cal O}^{(2)}$ has ten different SO(6) channels. Among them only three may contain unprotected superconformal primary operators: $[0,0,0]$, $[0,2,0]$ and $[1,0,1]$. As we have already exploited the known information about the lowest dimensional operators in the singlet, now we move to the superconformal primaries of dimension 4  arising\footnote{There is only one $\half$-BPS operator of dimension 2 in the irrep $[0,2,0]$ and it does not produce any new restrictions on the four-point amplitude.} in the irrep $[0,2,0]$.

There are four superconformal primary operators of canonical dimension 4 which might contribute to the operator product expansion under study. One of them is the protected semi-short double-trace operator \cite{AFP}. The other three are long unprotected operators $\Theta,\Theta_{\pm}$ whose one-loop anomalous dimensions were found in Ref. \cite{BERS} (see also \cite{Beisert}):
\bea
\gamma_{\Theta}=3\lambda\, , ~~~~~
\gamma_{\Theta_{\pm}}=\frac{1}{2}(5\pm\sqrt{5})\lambda \, .
\eea
The protected operator as well as $\Theta$ are double-traces, therefore we do not expect them to contribute to the four-point amplitude in the large $N$-limit. Denoting by $B_{\pm}$ the normalizations of the CPWAs of the operators $\Theta_{\pm}$ and proceeding in the same manner as before, we find the following system of equations:
\bea
&&B_{+}+B_{-}=\frac{147}{4N^2} \, ,\\
&&
B_{+}\gamma_{\Theta_+}+B_{-}\gamma_{\Theta_-}=\frac{105\lambda}{2N^2} \, ,\\
&&
B_{+}\gamma_{\Theta_+}^2+B_{-}\gamma_{\Theta_-}^2=\frac{35\lambda^2}{4 N^2}(9+2\mu) \, .
\eea
The first two equations are used to find
\bea
B_{\pm}= \frac{21}{8N^2}(7\mp 3\sqrt{5})\, .
\eea
The normalization constants are positive, as they should (the squares of the normalization coefficients of the three-point functions). Substituting these values in the third equation we find $\mu=0$.

Thus, by exploiting the information about the one-loop anomalous dimensions of some superconformal primaries, we have completely determined the freedom in the most general Ansatz (\ref{2loop3}):
\bea
\label{operestr}
\varrho=\mu=0\, , ~~~~~\eta=\nu=\frac{3^2}{4} \, .
\eea
One can easily see that with these values the amplitude (\ref{2loop3}) indeed coincides with the function $\cF^{(3)}_+$ found in the previous section.

Finally, we remark that 
in Ref. \cite{BKS} the two-loop
anomalous dimensions for the operators $\Theta_{\pm}$ were obtained
\bea
\label{2ldim}
\gamma_{\pm}=-\frac{17\pm 5\sqrt{5}}{8}\lambda^2 \, .
\eea
Working out the logarithmic terms in the two-loop OPE implied by the four-point amplitude for $k=3$ we have checked that eq. (\ref{2ldim}) is indeed compatible with our findings provided the one-loop corrections to the normalization constants are
\bea
B_{\pm}=\la {\cal O}^{(3)}{\cal O}^{(3)} \Theta_{\pm} \ra^2=\frac{21}{8N^2}(7\mp 3\sqrt{5})\Big[1-
\frac{1}{20}(35\pm 3\sqrt{5})\lambda \Big] \, .
\eea

\subsection{Weight 4}
Now we obtain some restrictions on the four-point two-loop amplitude for the $\half$-BPS operators of weight 4 from the knowledge of the one-loop anomalous dimensions for the same operators of twist 2 and 4 that were already discussed in the previous section.

For $k=4$ we expect two functions ${\cal F}^{(4)}_{\pm}$ with the symmetry property
$$
{\cal F}^{(4)}_{\pm}(s,t)=1/t~{\cal F}^{(4)}_{\pm}(s/t,1/t)\, .
$$
The general Ans\"atze compatible with these symmetries are
\begin{eqnarray}\label{2loop4}
  \cF^{(4)}_{\pm}(s,t)&=&\frac{\lambda^2}{N^2}\cdot\Big[\frac{1}{4}\Big(\varrho_{\pm} s+\eta_{\pm}(t+1)\Big)[\Phi^{(1)}(s,t)]^2 \\
\nonumber
&+&
 \frac{\mu_{\pm}}{s}\Phi^{(2)}(t/s,1/s)+ \nu_{\pm}\Big(\Phi^{(2)}(s,t)
+ \frac{1}{t}\Phi^{(2)}(s/t,1/t)\Big)
 \Big] \, ,
\end{eqnarray}
i.e. each function depends on four unknown coefficients $\varrho,\eta,\mu$ 
and $\nu$.

\vskip 0.3cm
{\it Operators of twist 2.}
\vskip 0.3cm

This time the Konishi field (as well as its higher-spin cousins) imposes the following relations
\bea
\label{K4}
\eta_{+}=\nu_{+}=\frac{4^2}{4}=4\, ,
\eea
while the other coefficients remain invisible. Clearly, these values of $\eta_{+}$, $\nu_{+}$ coincide with those in eq. (\ref{minpl}) obtained by explicit diagrammatic computations.

\vskip 0.3cm
{\it Operators of twist 4 in the singlet channel.}
\vskip 0.3cm

A calculation similar to the case $k=3$ produces the following system (after disentangling the contributions of the superconformal descendants of the twist 2 fields):
\bea
\label{comp4}
\nonumber
&& A_++A_-=\frac{32}{15N^2}\, ,\\
\label{singlet4}
&& A_{+}\gamma_{\Sigma_+}+A_-\gamma_{\Sigma_-}=\frac{56}{15}
\frac{\lambda}{N^2}\, ,\\
\nonumber
&&A_{+}\gamma_{\Sigma_+}^2+A_-\gamma_{\Sigma_-}^2=
\frac{2}{3}\Big(3\varrho_+-\frac{6}{5}+\eta_-+2\mu_-+2\nu_-+\varrho_-
\Big)\frac{\lambda^2}{N^2}\, ,
\eea
where $\gamma_{\Sigma_{\pm}}$ are given by eqs. (\ref{dg}). Solving 
this system we find
\bea
\label{lm1}
\eta_-=12-3\varrho_+-2\mu_--2\nu_--\varrho_-\, .
\eea
This completes our consideration of the singlet channel.

\vskip 0.3cm
{\it Operators of twist 4 in the irrep $[0,2,0]$.}
\vskip 0.3cm

The occurrence of the superconformal primaries $\Theta_{\pm}$ in the  
double operator product expansion implied by the four-point amplitude 
leads to the following system of equations
\bea
&&B_++B_-=\frac{2744}{25N^2}\, ,
\non\\
&&
B_+\gamma_{\Theta_+}+B_-\gamma_{\Theta_-}=\frac{784\lambda}{5N^2} \, ,\\
\non
&&B_+\gamma_{\Theta_+}^2+B_-\gamma_{\Theta_-}^2=\frac{98\lambda^2}{5 N^2}
(\eta_-+2\mu_-+2\nu_-+\varrho_-) \, .
\eea
It allows us to find
\bea
\label{newcond}
\eta_-=12-2\mu_--2\nu_--\varrho_- \, .
\eea
Comparing the last formula to eq. (\ref{lm1}) gives
$
\varrho_+=0 $, 
which coincides with the corresponding value of the coefficient $\varrho_{+}$ 
in eq. (\ref{minpl}).

This exhausts the predictive power of the method. The remaining coefficients 
$\mu_{\pm}$, $\nu_-$ and $\rho_-$ can be found only from the diagrammatic 
computation. If we had the information about some other superconformal 
primaries appearing in the OPE for $k=3$ or $k=4$, we could use it to 
further refine our Ansatz.

Finally, according to formula (\ref{minpl}) we have
\bea
\mu_+=\eta_-=\nu_-=0\, , ~~~~\mu_-=\varrho_-=4  \, .
\eea
With these values eq. (\ref{newcond}) is trivially satisfied.

\section*{Acknowledgements} We are grateful to N. Beisert,
S. Frolov, S. Kovacs, H. Osborn, J. Plefka, Y. Stanev and
M. Staudacher for many useful discussions.
This work has been supported in
part by INFN, MURST and the European Commission RTN program
HPRN--CT--2000--00131, in which
G. A. is associated to
the University of Bonn, S. P. is associated to
the University of Padova and A. S. is associated to the University of
Torino. The work of G. A. has been also supported by RFBI grant N02-01-00695.

\newpage

\section*{Appendices}
\appendix
\setcounter{equation}0

\section{Harmonic analyticity constraints}

In this Appendix we develop a set of systematic rules to select
the relevant graphs contributing to a correlation function
and find consistency conditions for the combinatorial coefficients.
These rules are based on the general property of harmonic
analyticity satisfied by any correlation function of composite operators
in G--analytic harmonic superspace.

Harmonic analyticity for a composite operator ${\cal O}^k =
{\rm Tr}(\tilde q^n q^{k-n})$ means $D^{++}{\cal O}^k = 0$, where
$D^{++}$ is the harmonic derivative given in (\ref{D++}).
Correlation functions then satisfy
\begin{equation}\label{c1}
  D^{++}_i \langle {\cal O}^k(1) \cdots {\cal O}^k(n)\rangle =
0 \qquad i = 1, \cdots, n
\end{equation}
(modulo contact terms). 
At every order in perturbation theory we can implement this condition 
directly on Feynman graphs. As a consequence of the identity
\begin{equation}\label{c2}
  D^{++}_1 \langle \tilde q^+(1) q^+(2) \rangle = -\langle \tilde q^+(1)
q^+(2) \rangle \overleftarrow{D}^{++}_2  = \delta(1,2)
\end{equation}
where $\delta(1,2)$ is the delta function in the G-analytic harmonic
superspace,
every time the derivative hits a HM propagator shrinks it to a point.
In particular, when $D^{++}$ hits a free line we have a vanishing contribution
(we neglect contact terms), whereas when it hits a HM line involved in an
interaction with a gluon, the effect is to shrink the propagator
line to a point and move the internal gluon vertex to coincide with an
external vertex. A simple example is given in Fig. 9 where analyticity
implemented at the vertex 1 of diagram ${\rm a}_1$ produces the shrunk
configuration ${\rm a}_2$.

If the same shrunk configuration can be obtained when hitting
different graphs (but at the same point), we can say that all these graphs
conspire together to restore harmonic analyticity at that point.
Repeating this procedure, one eventually recovers the whole class of
graphs needed to form a harmonic analytic amplitude.

\vskip 20pt
\begin{figure}[ht]
\begin{center}
\includegraphics[width=0.7\textwidth]{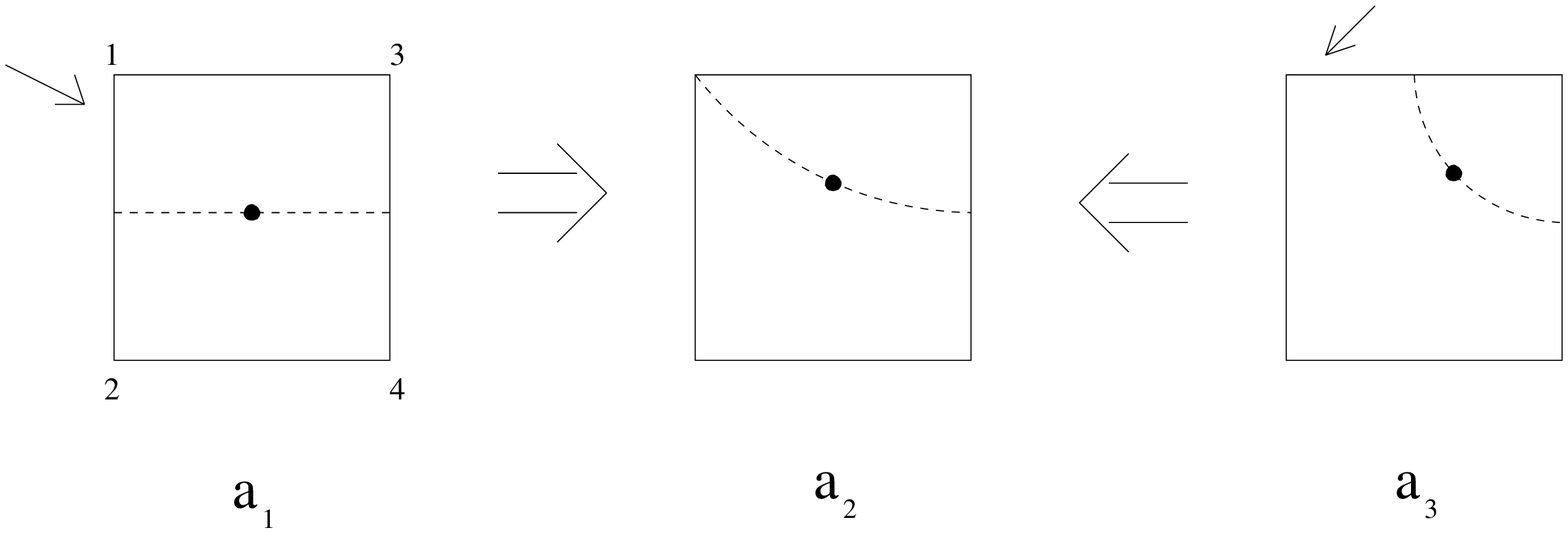}
\end{center}
\begin{center}
Figure 9. Example of an analyticity condition. The arrows point at
the propagator hit by $D^{++}_1$.
\end{center}
\end{figure}
\vskip 15pt

Again, we exemplify this by taking the simple case of Fig. 9.
There, the same shrunk configuration is obtained both
from diagrams ${\rm a}_1$ and ${\rm a}_3$.
Since these are the only two diagrams which
give the configuration ${\rm a}_2$ when hit at the point 1, the sum of their
contributions must result into an expression which is analytic in
the 1 variable. In fact, using eq. (\ref{T}) their contributions read 
\begin{eqnarray}
&& {\rm a}_1 \rightarrow
C_{a_1} [13][42] \left\{ [21^-]\rho_1^2 + [12^-]\rho_2^2 -
2(\rho_1 \rho_2) \right\}  \left\{ [43^-]\rho_3^2 + [34^-]\rho_4^2 -
2(\rho_3 \rho_4) \right\} \nonumber
\\
&& {\rm a}_3 \rightarrow
C_{a_3} [12][42] \left\{ [31^-]\rho_1^2 + [13^-]\rho_3^2 -
2(\rho_1 \rho_3) \right\}  \left\{ [43^-]\rho_3^2 + [34^-]\rho_4^2 -
2(\rho_3 \rho_4) \right\} \nonumber
\\
&& \quad
\label{c3}
\end{eqnarray}
where $C_{a_i}$ are the combinatorial factors (with possible signs included).
The two expressions are
non-analytic in the 1 variable because of their nontrivial dependence on
$u^-_1$.
The action of $D^{++}_1$ on the sum of the two contributions must
give zero, according to the condition (\ref{c1}).
Remembering that $D^{++}_1$ only acts on the $u_1^-$ harmonic converting it
into $u_1^+$, we obtain
\begin{equation}
\left( C_{a_1} + C_{a_3} \right) [42][13][21] \left\{ [43^-] \rho_1^2 \rho_3^2
+ [34^-] \rho_1^2 \rho_4^2 \right\} ~=~ 0
\end{equation}
which implies
\begin{equation}
C_{a_1} = - C_{a_3}
\label{c4}
\end{equation}
Using this condition, the sum of the two contributions in (\ref{c3}) reads
(we consider only the part depending on $1^-$)
\begin{equation}
[42] \left\{ [13][21^-] - [12][31^-] \right\} \left\{ [43^-] \rho_1^2 \rho_3^2
+ [34^-] \rho_1^2 \rho_4^2 \right\}
\end{equation}
The harmonic cyclic identity
\begin{equation}
[13][21^-] - [12][31^-] ~=~ [23]
\end{equation}
shows that the non-analytic contributions from the two diagrams
cancel against each other leaving an analytical result in the 1 variable.

As appears from this simple example, imposing analyticity conditions allows
to draw systematically all the diagrams which at a given order contribute
to the analytic final result for the correlation function and to find
nontrivial relations like (\ref{c4})
which constrain the combinatorial coefficients. These relations can be used
as a check when the coefficients are directly computed from the diagrams.

We notice that the final analytic result can be obtained by using a shortcut
procedure: Given the sum of the two $1^-$ dependent contributions 
in (\ref{c3}) we make the harmonics identification $1 \equiv 2$. As a
consequence, the contribution from a$_3$ vanishes whereas a$_1$ gives
\begin{equation}
C_{a_1} [42] [13][21^-] \left\{ [43^-] \rho_1^2 \rho_3^2
+ [34^-] \rho_1^2 \rho_4^2 \right\}
\stackrel{1\equiv 2}{\Longrightarrow}
C_{a_1} [42][23] \left\{ [43^-] \rho_1^2 \rho_3^2
+ [34^-] \rho_1^2 \rho_4^2 \right\}
\end{equation}
which is the correct result.
The identification of the harmonics we chose is not the only 
possible one. We could have made the alternative identification $1 \equiv 3$.
In this case the contribution from diagram a$_1$ vanishes whereas from
diagram a$_3$ we obtain 
\begin{equation}
C_{a_3} [12] [42][31^-] \left\{ [43^-] \rho_1^2 \rho_3^2
+ [34^-] \rho_1^2 \rho_4^2 \right\}  
\stackrel{1\equiv 2}{\Longrightarrow}
- C_{a_3} [23][42] \left\{ [43^-] \rho_1^2 \rho_3^2
+ [34^-] \rho_1^2 \rho_4^2 \right\}
\end{equation}
which again is the correct result.

\vskip 10pt
We now consider the cases of our interest.
We start by studying the analyticity conditions for the set of diagrams
of Fig. 8 in the main text.
They have in common two free lines which give the analytic structure
$[12][43]$ and a ``base'' $T_{245}T_{246}$:
\bea \nonumber
T_{245}T_{246}&=&[24^-][42^-](\rho^2_4\sigma^2_2 + \rho^2_2\sigma^2_4)+4(\r_2\r_4)(\sigma_2\sigma_4) \\
\label{TxT} &=& \tau_{24} -[24][2^-4^-](\rho^2_4\sigma^2_2 +
\rho^2_2\sigma^2_4) \, . 
\eea   
It is easy to see that leaving aside the base as well as the free
propagators leads to the following harmonic and Grassmann structures:
\begin{eqnarray}
 (8a):&& [13]^2 [T_{125}T_{436} + 5\leftrightarrow6] =  [13]^2[21^-][43^-]
(\rho^2_1\sigma^2_3 + \rho^2_3\sigma^2_1)\nonumber\\
 (8b):&& [13][43][T_{125}T_{136} + 5\leftrightarrow6] =  [13][43][21^-]
[13^-](\rho^2_1\sigma^2_3 + \rho^2_3\sigma^2_1)\nonumber\\
 (8c):&& [13][12][T_{435}T_{136} + 5\leftrightarrow6] =  [13][12][43^-]
[31^-](\rho^2_1\sigma^2_3 + \rho^2_3\sigma^2_1)\nonumber\\
 (8d):&& [12][43][T_{135}T_{136} + 5\leftrightarrow6] =  2[12][43]\left\{-[13]
[1^-3^-](\rho^2_1\sigma^2_3 + \rho^2_3\sigma^2_1) + \tau_{13}\right\}
\nonumber\\
\end{eqnarray}
The factor 2 in the graph $(8d)$ is due to the symmetrization
$5\leftrightarrow6$. Note that this is the only graph from this set which
gives an analytic contribution proportional to $\tau_{13}$. The other
Grassmann term $\rho^2_1\sigma^2_3 + \rho^2_3\sigma^2_1$ always appears
with non-analytic harmonic factors. Collecting all such terms we find
\begin{equation}\label{c5}
  [13]\left\{ C_a [12][43^-][31^-]+ C_b [13][21^-][43^-] +
C_c [43][21^-][13^-] - 2C_d [12][43][1^-3^-] \right\}
\end{equation}
where we have introduced combinatorial factors $C_a,\cdots,C_d$.
We want to achieve harmonic analyticity at points 1 and 3.
According to the rule described above, we hit the graphs with
harmonic derivatives $D^{++}_1$ and $D^{++}_3$.

We begin by applying $D^{++}_1$ to the expression (\ref{c5}) to obtain
\begin{equation}\label{c6}
  [13][12] \left\{ (-C_b-2C_d)[43][13^-] + (C_a+C_c)[43^-][31] \right\} = 0
\end{equation}
Since the two harmonic structures within the brackets are linearly independent,
we have the following two conditions
\begin{equation}\label{c7}
  C_b = -2C_d \qquad C_a=-C_c
\end{equation}
Similarly, hitting (\ref{c5}) with $D^{++}_3$  we obtain
\begin{equation}\label{c8}
  [13][43] \left\{ (C_a+C_b)[21^-][13] + (C_c+2C_d) [12][31^-] \right\} = 0
\end{equation}
which implies
\begin{equation}\label{c9}
  C_c = -2C_d  \qquad C_a=-C_b
\end{equation}
The last set of conditions is consistent with (\ref{c7}) if
$C_b = C_c$. This was somehow expected since graphs 8b) and 8c) only
differ by a reflection ($1 \leftrightarrow 3$, $2 \leftrightarrow 4$)
which should not affect the combinatorial factor.

We can solve the constraints (\ref{c7}) ending with only one arbitrary
coefficient, e.g. $C_d$. With this choice for the coefficients the
expression (\ref{c5})  becomes harmonic analytic with the help of the
identity
\begin{equation}
\label{ide}
  [43][21^-][13^-] + [12][43^-][31^-] - [13][21^-][43^-] + [12][43][1^-3^-]
= [42]
\end{equation}

Finally, taking all this into account, attaching the ``base'' and
restoring the missing space-time factors, we find the total
contribution of the graphs in Fig. 8 to be
\begin{eqnarray}
\mbox{Total Fig. 8:}  && \frac{2C_d [12][43]}{x^4_{12}x^4_{13}x^4_{24}x^4_{34}}
\left\{ - [13][42] (\rho_1^2 \sigma_3^2 + \rho_3^2 \sigma_1^2) + 
[12][43] \tau_{13} \right\} 
\nonumber\\
  && \times \left\{ -[24][2^-4^-](\rho^2_4\sigma^2_2 + \rho^2_2\sigma^2_4) +
\tau_{24} \right\} \label{tot7}
\end{eqnarray}
We see that analyticity at points 1 and 3 has indeed been achieved
but not at points 2 and 4, of course.

As in the previous example, the correct result can be obtained 
by using harmonics identification. In fact, pulling out the free
line contribution $[12][43]$, we can identify the harmonics 
$1 \equiv 2$ and $3 \equiv 4$. As a consequence, in eq. (\ref{c5})
the contributions from diagrams a,c,d vanish and we are left with 
\bea
C_b [13]^2[21^-][43^-](\rho^2_1\sigma^2_3 + \rho^2_3\sigma^2_1)
\stackrel{1\equiv 2,3\equiv 4}{\Longrightarrow}
C_b [13]^2(\rho^2_1\sigma^2_3 + \rho^2_3\sigma^2_1) \, . 
\eea 
Since the final result must be analytic, we can find an analytic term
which under the above identification brings to the same expression 
\bea
-C_b [13][42](\rho^2_1\sigma^2_3 + \rho^2_3\sigma^2_1)
\stackrel{1\equiv 2,3\equiv 4}{\Longrightarrow}
C_b [13]^2(\rho^2_1\sigma^2_3 + \rho^2_3\sigma^2_1) \, . 
\eea 
Therefore this must be the correct term which appears in the final
result and it is indeed what we found in (\ref{tot7}).

We notice that using this shortcut we can easily figure out the
complete analytic structure for the sum of diagrams in fig. 8. In fact,
analyticity at points $2$ and $4$, which should be restored by adding 
another set of partner graphs, can rather be obtained by identifying
$2\equiv 4$.  This identification leaves the single term $\tau_{24}$ in the 
bottom part of eq. (\ref{tot7}) which becomes completely analytic. 

\vskip 20pt
\begin{figure}[ht]
\begin{center}
\includegraphics[width=0.7\textwidth]{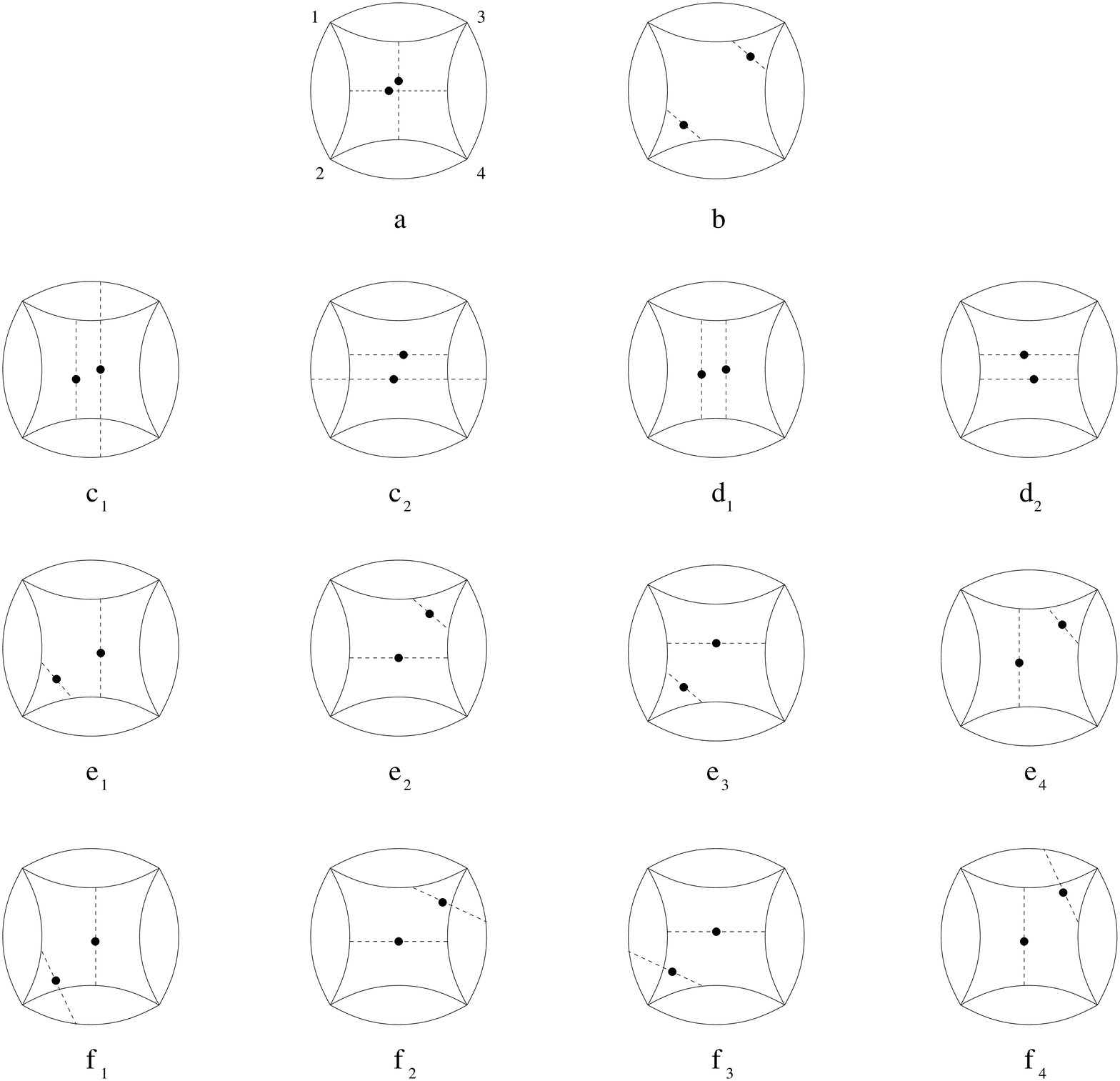}
\end{center}
\begin{center}
Figure 10. Graphs conspiring to harmonic analyticity at points 1 and 4.
\end{center}
\end{figure}
\vskip 15pt

We now consider the more complicated situation of graphs in Fig. 10.
By imposing analyticity at points 1 and 4 we
can find consistency conditions between the coefficients of the
graphs 10c) which have been checked independently in Section 6,
and the coefficients 10a) and 10b) which don't have any independent
check. These conditions will necessarily involve new graphs
contributing only to non-analytic structures, necessary in order
to cancel similar terms coming from the graphs 10a),
10b) and 10c).

We start by considering the graph a) and apply the outlined procedure in
order to achieve harmonic analyticity at points 1 and 4. It is
easy to realize that in the large $N$ limit, i.e. neglecting non-planar
graphs, the set of conspiring graphs is
just the one showed in Fig. 10.

As in the previous example, we
write the harmonic and Grassmann factors coming from these
graphs:

\begin{eqnarray}
(a):&& [12][13][42][43][T_{135}T_{126}T_{425}T_{436} +
5\leftrightarrow6] \longrightarrow\nonumber\\
&& [12][13][42][43]
\Big\{\big[[31^-][12^-][24^-][43^-]+[13^-][21^-][42^-][34^-]\big]\nonumber\\
&&\qquad\qquad\qquad\quad
\big[\rho^2_1\sigma^2_2\sigma^2_3\rho^2_4+\sigma^2_1\rho^2_2\rho^2_3
\sigma^2_4\big]\Big\}\nonumber\\
&&\nonumber\\
(b):&& [12][13][42][43][T_{136}T_{125}T_{425}T_{436} +
5\leftrightarrow6]\longrightarrow\nonumber\\
&& [12][13][42][43]
\Big\{[31^-][12^-][24^-][43^-]\big[\rho^2_1\sigma^2_2\rho^2_3\sigma^2_4+
\sigma^2_1\rho^2_2\sigma^2_3\rho^2_4\big]\nonumber\\
&&\qquad+[13^-][21^-][42^-][34^-]\big[\sigma^2_1\sigma^2_2\rho^2_3\rho^2_4
+\rho^2_1\rho^2_2\sigma^2_3\sigma^2_4\big]\Big\}\nonumber
\end{eqnarray}
\begin{eqnarray}
(c_1):&& [12]^2[43]^2[T_{135}T_{136}T_{425}T_{426} +
5\leftrightarrow6]\longrightarrow\nonumber\\
&& 2[12]^2[43]^2 \big\{-[13][1^-3^-](\rho^2_1\sigma^2_3+\rho^2_3\sigma^2_1)
+ \tau_{13} \big\} \left\{
-[24][2^-4^-](\rho^2_2\sigma^2_4+\rho^2_4\sigma^2_2)+\tau_{24}\right\}
\nonumber\\
(c_2):&& \hbox{same as } (c_1) \hbox{ with } 2\leftrightarrow 3 \hbox{ or }
1\leftrightarrow 4\nonumber\\
&&\nonumber\\
(d_1):&& [12]^2[13][42][43]^2[(TT)_{1563}(TT)_{4652} +
5\leftrightarrow6] \longrightarrow\nonumber\\
&&[12]^2[13][42][43]^2[1^-3^-][4^-2^-]\big(\rho^2_1\rho^2_2\sigma^2_3\sigma^2_4
+\sigma^2_1\sigma^2_2\rho^2_3\rho^2_4\big)
\nonumber\\
(d_2):&& \hbox{same as } (d_1) \hbox{ with } 2\leftrightarrow 3 \hbox{ or }
1\leftrightarrow 4\nonumber\\
&&\nonumber\\
(e_1):&& [12][13][42][43]^2[T_{135}T_{126}(TT)_{4562} + 5\leftrightarrow6]
\longrightarrow\nonumber\\
&&[12][13][42][43]^2[2^-4^-][13^-][21^-]\big(\sigma^2_1\sigma^2_2\rho^2_3
\rho^2_4+\rho^2_1\rho^2_2\sigma^2_3\sigma^2_4\big)
\nonumber\\
(e_2):&& \hbox{same as } (e_1) \hbox{ with } 2\leftrightarrow 3\nonumber\\
(e_3):&& \hbox{same as } (e_1) \hbox{ with } 1\leftrightarrow 4\nonumber\\
(e_4):&& \hbox{same as } (e_1) \hbox{ with } 2\leftrightarrow 3 \hbox{ and }
1\leftrightarrow 4\nonumber\\
&&\nonumber\\
(f_1):&& [12][13][43]^2[T_{135}T_{126}T_{425}T_{426}+ 5\leftrightarrow6]
\longrightarrow\nonumber\\
&&
[12][13][43]^2[13^-][21^-]\big(\sigma^2_1\rho^2_3+\rho^2_1\sigma^2_3\big)
\left\{-[24][2^-4^-](\rho^2_2\sigma^2_4+\rho^2_4\sigma^2_2)+\tau_{24}\right\} \nonumber\\
(f_2):&& \hbox{same as } (f_1) \hbox{ with } 2\leftrightarrow 3\nonumber\\
(f_3):&& \hbox{same as } (f_1) \hbox{ with } 1\leftrightarrow 4\nonumber\\
(f_4):&& \hbox{same as } (f_1) \hbox{ with } 2\leftrightarrow 3 \hbox{ and }
1\leftrightarrow 4
\label{cont10}
\end{eqnarray}
We concentrate on the contributions to the three independent
structures not containing $\tau$ factors
\begin{eqnarray}
U&\equiv& \rho^2_1\rho^2_2\sigma^2_3\sigma^2_4 +
\sigma^2_1\sigma^2_2\rho^2_3\rho^2_4\nonumber\\
V&\equiv& \rho^2_1\sigma^2_2\rho^2_3\sigma^2_4 +
\sigma^2_1\rho^2_2\sigma^2_3\rho^2_4\nonumber\\
Z&\equiv& \rho^2_1\sigma^2_2\sigma^2_3\rho^2_4 +
\sigma^2_1\rho^2_2\rho^2_3\sigma^2_4
\end{eqnarray}
They are
\begin{eqnarray}
(a):&& C_a[12][13][42][43]\Big\{[31^-][12^-][24^-][43^-]+[13^-][21^-]
[42^-][34^-]\Big\} Z\nonumber\\
&&\nonumber\\
(b):&& C_b[12][13][42][43]\Big\{ [31^-][12^-][24^-][43^-] V +[13^-][21^-]
[42^-][34^-] U\Big\} \nonumber\\
&&\nonumber\\
(c_1):&& 2C_c[12]^2[43]^2[13][1^-3^-][24][2^-4^-](U+Z) \nonumber\\
(c_2):&& \hbox{same as } (c_1) \hbox{ with } 2\leftrightarrow 3 \hbox{ or }
1\leftrightarrow 4\nonumber\\
&&\nonumber\\
(d_1):&& C_d[12]^2[13][42][43]^2[1^-3^-][4^-2^-] U\nonumber\\
(d_2):&& \hbox{same as } (d_1) \hbox{ with } 2\leftrightarrow 3 \hbox{ or }
1\leftrightarrow 4\nonumber\\
&&\nonumber
\end{eqnarray}

\begin{eqnarray}
(e_1):&& C_e[12][13][42][43]^2[2^-4^-][13^-][21^-] U\nonumber\\
(e_2):&& \hbox{same as } (e_1) \hbox{ with } 2\leftrightarrow 3\nonumber\\
(e_3):&& \hbox{same as } (e_1) \hbox{ with } 1\leftrightarrow 4\nonumber\\
(e_4):&& \hbox{same as } (e_1) \hbox{ with } 2\leftrightarrow 3 \hbox{ and }
1\leftrightarrow 4\nonumber\\
&&\nonumber\\
(f_1):&& C_f[12][13][43]^2[13^-][21^-][42][2^-4^-] (U+Z) \nonumber\\
(f_2):&& \hbox{same as } (f_1) \hbox{ with } 2\leftrightarrow 3\nonumber\\
(f_3):&& \hbox{same as } (f_1) \hbox{ with } 1\leftrightarrow 4\nonumber\\
(f_4):&& \hbox{same as } (f_1) \hbox{ with } 2\leftrightarrow 3 \hbox{ and }
1\leftrightarrow 4
\label{contrgraphs}
\end{eqnarray}
We have assumed that diagrams which differ by the exchange
$2\leftrightarrow 3$ or/and $1\leftrightarrow 4$ have the same combinatorial
factor.

By acting with $D_1^{++}$ and setting to zero the independent
structure proportional to $U$ we get
\begin{eqnarray}
&&[12][13][42][43]\Big\{[12][43][13^-][2^-4^-]\big(-2C_c-C_d-C_e-C_f\big)
\nonumber\\
&&\qquad\qquad\qquad\quad
+[12][34^-][13^-][42^-]\big( -C_b-C_e-C_f\big) \Big\}=0
\label{condU}
\end{eqnarray}
Doing the same for the structure proportional to $V$ gives an
equation like (\ref{condU}) with $2\leftrightarrow 3$, then it
will not impose any new condition on the coefficients. The part
proportional to $Z$ gives instead
\begin{eqnarray}
&&[12][13][42][43]\Big\{[12][43][13^-][2^-4^-]\big(-2C_c-C_f\big)\nonumber\\
&&\qquad\qquad\qquad\quad
+[12][34^-][13^-][42^-]\big(-C_a-C_f\big) \Big\} + 2\leftrightarrow
3=0 \label{condZ}
\end{eqnarray}
By now setting to zero the two linearly independent analytic
structure appearing in (\ref{condU}) and (\ref{condZ}) we get the
following conditions relating the coefficients of the graphs of
Fig.10
\begin{eqnarray}
C_a &=& 2C_c =-C_f\nonumber\\
C_b &=& C_a-C_e =C_a+C_d \label{condfig10}
\end{eqnarray}
We notice that the condition $C_f= -2C_c$ has been already obtained
in (\ref{c9}) (diagrams 10c) and 10f) are the same as 8d) and 8c),
respectively).

It's now easy to see that there are no extra conditions coming
from the terms containing one $\tau$ factor. The same is true
if we impose the analyticity condition at point 4.
So we conclude that (\ref{condfig10}) is the full set of conditions
imposed by the requirement of partial analyticity at points 1 and
4 on the graphs in Fig.10.

Using the relations (\ref{condfig10}) one can check that the total
contribution obtained by summing the expressions in (\ref{cont10}) ends up
to be analytic in the $1$ and $3$ variables. 

Again, as in the previous examples, the final answer can be obtained by
suitable identification of the harmonics.

To summarize, what we have leaned from the examples described above is
the following: On one hand, imposing harmonic analyticity on the final 
result allows to find consistency conditions among the coefficients of
different sets of graphs. On the other hand, once we give analyticity for 
granted, we can bypass the whole procedure of evaluating all diagrams 
conspiring to a final analytic result by performing harmonics 
identification, so drastically reducing the number of graphs to be computed.

\section{Sample calculations: $k=3$ and $k=4$}

In this Appendix we describe in detail the calculation of the $A_m$ functions
(see eq. (\ref{Am})) for the cases $k=3$ and $k=4$. In particular, we show how the procedure of identifying the harmonics can be used both to unambiguously select the diagrams which contribute to a given function and 
to drastically reduce the number of diagrams one needs compute.

\subsection{The case $k=3$}

The expected form of the complete amplitude for $k=3$ case is:
\begin{equation}\label{complam}
 \langle Q^{(3)}|{\cal L}|{\cal L}\rangle_{\theta_{5,6}=0} 
\sim \prod_{r=1}^4 (\theta^+_r)^2  
\left[\frac{[13][42]}{x^2_{13}x^2_{42}} A_0(x)  + \frac{[12][43]}{x^2_{12}
x^2_{43}} A_1(x)  \right]
\end{equation}
As discussed in the main text, the two functions $A_0, A_1$ are related by 
crossing symmetry, and we need compute only one of them. 
To calculate for example $A_0$, it is sufficient to look at contributions
proportional to the harmonic structure $[13][42]$ in (\ref{complam}). 
They can be unambiguously selected by identifying the harmonics 
$1 \equiv 2$ and $3 \equiv 4$ in (\ref{complam}) 
\begin{equation}\label{compind}
\langle Q^{(3)}|{\cal L}|{\cal L}\rangle_{\theta_{5,6}=0}  
\sim \prod_{r=1}^4(\theta^+_r)^2
\left[ \frac{[13]^2}{x^2_{13}x^2_{42}} A_0(x)  
\right]
\end{equation}
When looking at perturbative contributions at two loops, the 
harmonics identification implies the vanishing of all graphs containing:

(i)   at least one free line $1\rightarrow2$ or $3\leftarrow4$

(ii)  the blocks $TT_{1562}$ or $TT_{3564}$

(iii) the blocks $T_{125}T_{125}$ and alike 

\noindent
Discarding also the graphs which vanish by simple theta counting (see the general
discussion in Sect. 5), the relevant surviving diagrams are then shown in 
Fig. 11. Upon the above harmonic identification, they already contain the analytic
factor $[13]^2$ required 
in eq. (\ref{compind}). Since the coefficient function $A_0$ is independent
of the harmonics, once we have pulled $[13]^2$ out, we can go a step further 
and identify the harmonics $1\equiv3$ in the rest of the expression.

As a consequence, the contributions from graphs 11c and 11d drop out.  
In the rest the harmonic dependence reduces to a common factor and, neglecting the space-time structure, the final contribution 
of each diagram is given by its combinatorics
\begin{eqnarray}
&& 2 C^{(11)}_a = 2 \times 3^4 N^6
\nonumber \\
&& C_b^{(11)} = C_e^{(11)} = C_f^{(11)} = 3^4 N^6
\nonumber \\
&& C_{g_1}^{(11)} = C_{g_2}^{(11)} = 2 \times 3^4 N^6
\nonumber \\
&& 2 C_h^{(11)} = 3^4 N^6
\end{eqnarray}

\vskip 15pt
\noindent
\begin{minipage}{\textwidth}
\begin{center}
\includegraphics[width=0.60\textwidth]{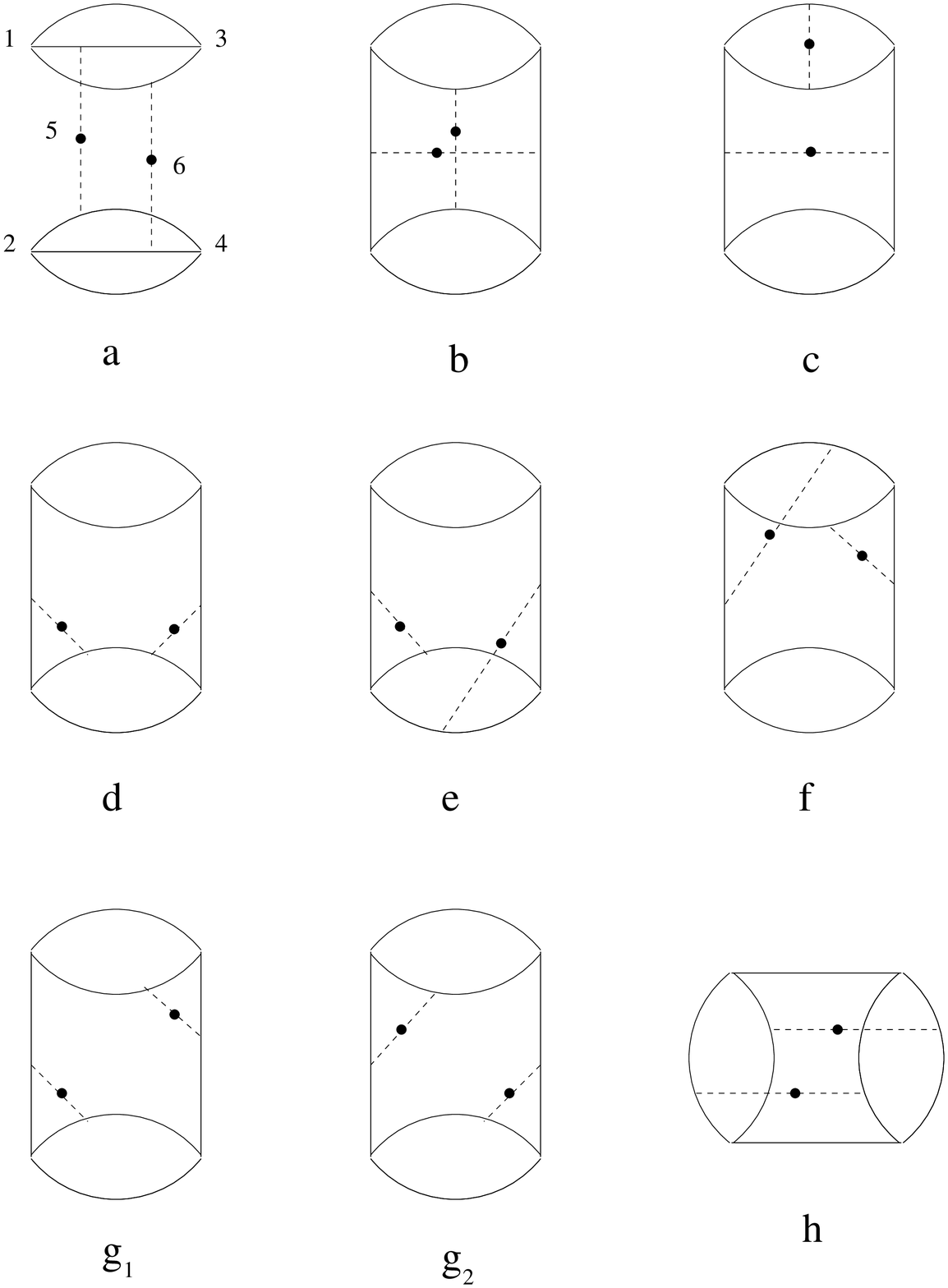}
\end{center}
\begin{center}
Figure 11. Relevant graphs for the $k=3$ case.
\end{center}
\end{minipage}

\vskip 20pt

The extra factor 2 in front of $C_a^{(11)}$ and $C_h^{(11)}$ is due
to the symmetry of the corresponding diagrams under the exchange $5
\leftrightarrow 6$.

In this simple example it is already clear how one can exploit the condition
of harmonic analyticity of the final result. It is in fact this condition 
which allowed us to make the identification $1 \equiv 3$, thus reducing the
number of Feynman diagrams to be computed. In a complete calculation performed
without identifying the harmonics, the diagrams which vanished under this identification would contribute only to cancel non-analytic terms coming
from the rest of the diagrams.

\subsection{The case $k=4$}

We now consider the more complicated case $k=4$. 
There we have
\begin{eqnarray}
  \langle Q^{(4)}|{\cal L}|{\cal L}
\rangle_{\theta_{5,6}=0} \ &\sim &\prod_{r=1}^4 (\theta^+_r)^2
 \label{complam4}\\   
&&\hspace{-1cm} \times \left[\left(\frac{[13][42]}{x^2_{13}x^2_{42}}\right)^2 A_0(x)  + 
\left(\frac{[13][42]}{x^2_{13}x^2_{42}}\frac{[12][43]}{x^2_{12}x^2_{43}}\right)
 A_1(x) + \left(\frac{[12][43]}{x^2_{12}x^2_{43}}\right)^2 A_2(x)  \right] \, .
\nonumber
\end{eqnarray}
This correlator is crossing symmetric under the exchange  
$2\leftrightarrow3$ which relates the coefficient functions $A_0$ and $A_2$ 
to each other,  while $A_1$ is crossing symmetric and independent. 
Therefore, we need to perform the calculation in two independent channels.

To select the relevant graphs which contribute in the two channels
we observe that in the $k=4$ case we have 8 HM lines and 2 gluons, which can 
at most connect 4 HM lines. Therefore, at least 4 HM lines remain free. 
If these free lines form a free corner, i.e. they all come out from the same 
vertex, we know that such graphs vanish because of 
theta counting (see the general discussion in sect. 5) . It follows that
the free lines have to form at least one disconnected pair. 

If the free pair is $[13][42]$, such graphs can contribute to the functions 
$A_0,A_1$ in eq. (\ref{complam4}) but not to $A_2$. Then we can write down 
the contribution of all such graphs with a free pair $[13][42]$ in the form 
\begin{equation}\label{pair13}
  \frac{[13][42]}{x^2_{13}x^2_{42}} \left[\frac{[13][42]}{x^2_{13}x^2_{42}} 
A_0(x) + \frac{[12][43]}{x^2_{12}x^2_{43}} A_1(x)  \right]
\end{equation}
Pulling out the factor $\frac{[13][42]}{x^2_{13}x^2_{42}}$ is equivalent to 
removing the corresponding free lines from the graphs, which results in a 
configuration with $k=3$. Then we can go a step further and identify the 
harmonics pairwise within the brackets. If we identify 
$1\equiv2$, $3\equiv4$ and the graph does not vanish, then it contributes 
to $A_0$. If instead we identify $1\equiv3$, $2\equiv4$ and the graph does 
not vanish, then it contributes to $A_1$. Note that it may happen that the 
graph vanishes under each of these identifications, then it should be 
discarded.

Alternatively, if the free pair is $[12][34]$, eq. (\ref{pair13}) is replaced 
by
\begin{equation}\label{pair12}
  \frac{[12][43]}{x^2_{12}x^2_{43}} \left[\frac{[13][42]}{x^2_{13}x^2_{42}} 
A_1(x) + \frac{[12][43]}{x^2_{12}x^2_{43}} A_2(x)  \right]
\end{equation}  
Once again, removing the free lines we obtain a $k=3$ configuration in which 
we identify the harmonics pairwise within the brackets. If we identify 
$1\equiv2$, $3\equiv4$ and the graph does not vanish, then it contributes to 
$A_1$. If instead we identify $1\equiv3$, $2\equiv4$ and the graph does not 
vanish, then it contributes to $A_2$.   

We now use this strategy to select the relevant diagrams which eventually
contribute to the coefficient functions. 
We carefully draw all the diagrams which are
nonvanishing for theta counting and select the ones which survive
under one of the harmonics identifications described above. We are then led 
to the set of graphs on Fig. 12 where, in order to reduce the number of
diagrams, we have indicated with thick lines all free HM propagators, whereas
thin lines are interacting (they contain an interaction vertex and
gluons connect these vertices in all possible ways). 
All the graphs have 4 free lines and they have been organized according to 
the number of free lines coming out of a single point.
We have not drawn those diagrams that vanish
after harmonics identification (the analog of graphs 11c and 11d of
the previous example). As already explained,
in an exhaustive calculation done without identifying harmonics 
these diagrams would conspire to cancel nonanalytic contributions from the
rest. 

\vskip 15pt
\noindent
\begin{minipage}{\textwidth}
\begin{center}
\includegraphics[width=0.70\textwidth]{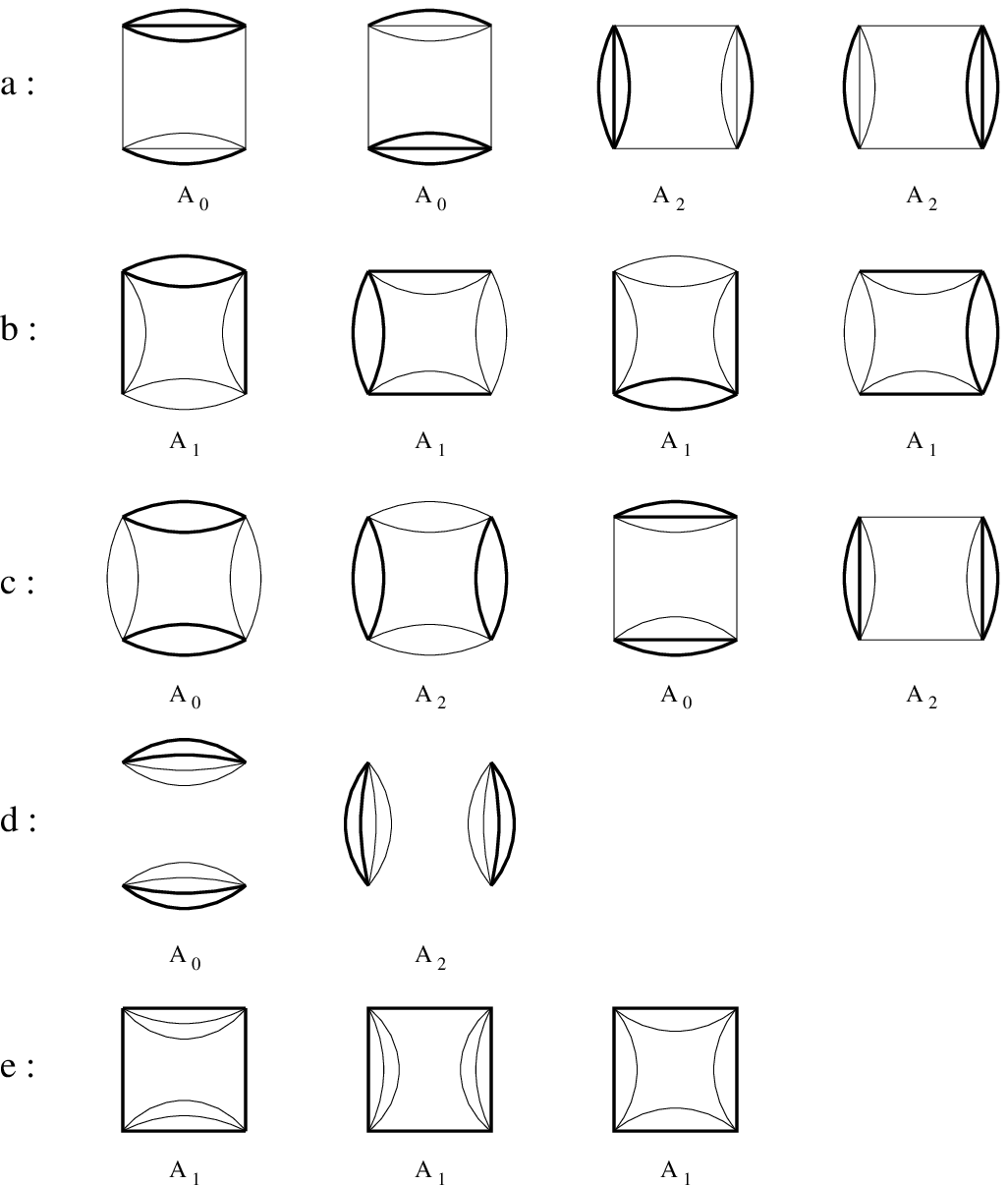}
\end{center}
\begin{center}
Figure 12. Relevant graphs for $k=4$. The thick lines are free HM propagators
whereas thin lines contain vertices of interaction with gluons. 
\end{center}
\end{minipage}

\vskip 20pt

We now apply the harmonics identification procedure to figure out which
diagrams contribute to which function. 
Let us analyze in detail the graphs in Fig 12a. 
The first kind of graphs contain a free pair $[13][42]$, so they belong 
to the type of eq. (\ref{pair13}). According to the rule above, we pull out 
the 
free factor and then we identify the harmonics pairwise. The identification 
$1\equiv3$, $2\equiv4$ annihilates the graphs, so they cannot contribute to 
$A_1$. On the contrary, the graphs can survive the  identification 
$1\equiv2$, $3\equiv4$, so they contribute to $A_0$. If we now draw the 
gluons, we see that only one configuration is allowed and the graph is 
reduced to 11e from the case $k=3$. 
The same argument applies to the second kind of graphs 12a, 
reducing them to a single structure like 11f. 
Similarly, the third and the fourth graphs 12a contain  a free pair 
$[12][43]$, so they belong to the type of eq. (\ref{pair12}). 
According to the rule above, we pull out the free factor and then we 
identify the harmonics pairwise. The identification $1\equiv2$, $3\equiv4$ 
annihilates the graphs, so they cannot contribute to $A_1$. 
On the contrary, the graphs survive the  identification 
$1\equiv3$, $2\equiv4$, so they contribute to $A_2$. 
If we now draw the gluons, we see that these graphs are reduced to 11e, 11f 
rotated by $90^o$. 

Analyzing all the graphs along the same lines we eventually identify the
coefficient functions they contribute to. The complete identification is
indicated in Fig. 12. Drawing the gluon lines brings us to the diagrams in
Fig. 13 for $A_0$ and Fig. 14 for $A_1$. 

\vskip 15pt
\noindent
\begin{minipage}{\textwidth}
\begin{center}
\includegraphics[width=0.70\textwidth]{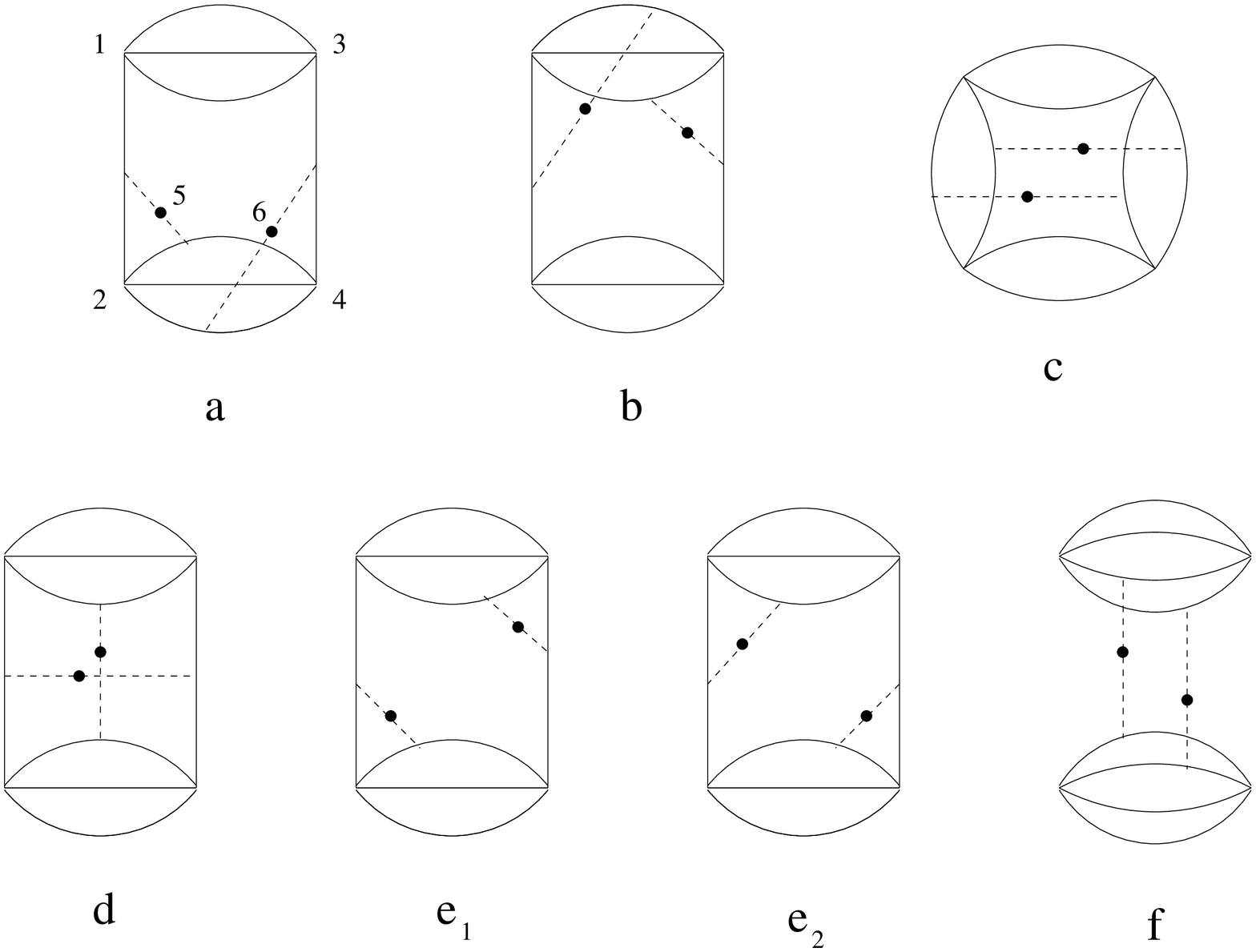}
\end{center}
\begin{center}
Figure 13. Diagrams contributing to the $A_0$ function for $k=4$ case. 
\end{center}
\end{minipage}
\vskip 20pt

A couple of
comments are now in order. First of all we notice that the 
configurations of gluon lines appearing in Figs.13 and 14 are the 
same as the ones we have already selected for $k=3$. Moreover, a
diagram with a given configuration of interacting lines can contribute to
different functions according to the structure of free propagators which
dress it. In fact, all the allowed configurations of interacting lines
are present for both functions. 

Now, computing the contribution of each graph is simply a matter of 
combinatorics. At large $N$, neglecting the space-time structure, for
the $A_0$ function we obtain
\begin{eqnarray}
&& C^{(13)}_a = C_b^{(13)} = 4^4 N^8
\nonumber \\
&& 2C_c^{(13)} = 4^4 N^8
\nonumber \\
&& C_d^{(13)} = 4^4 N^8
\nonumber \\
&& C^{(13)}_{e_1} = C_{e_2}^{(13)} = 2 \times 4^4 N^8
\nonumber \\
&& 2C_f^{(13)} = 2 \times 4^4 N^8
\end{eqnarray}
whereas for $A_1$
\begin{eqnarray}
&& C^{(14)}_a = C_b^{(14)} = C_c^{(14)} =C_d^{(14)} = 4^4 N^8
\nonumber \\
&& 2C_e^{(14)} = 4^4 N^8
\nonumber \\
&& 2C_f^{(14)} = 4^4 N^8
\nonumber \\
&& C_g^{(14)} = 4^4 N^8
\nonumber \\
&& C^{(14)}_{h_1} = C_{h_2}^{(14)} = 2 \times 4^4 N^8
\end{eqnarray}
It is important to notice that, at large $N$, the combinatorial factors
only depend on the structure of the interacting lines: Diagrams with 
the same interactions but different configurations of free propagators
have the same coefficient.

\vskip 15pt
\noindent
\begin{minipage}{\textwidth}
\begin{center}
\includegraphics[width=0.60\textwidth]{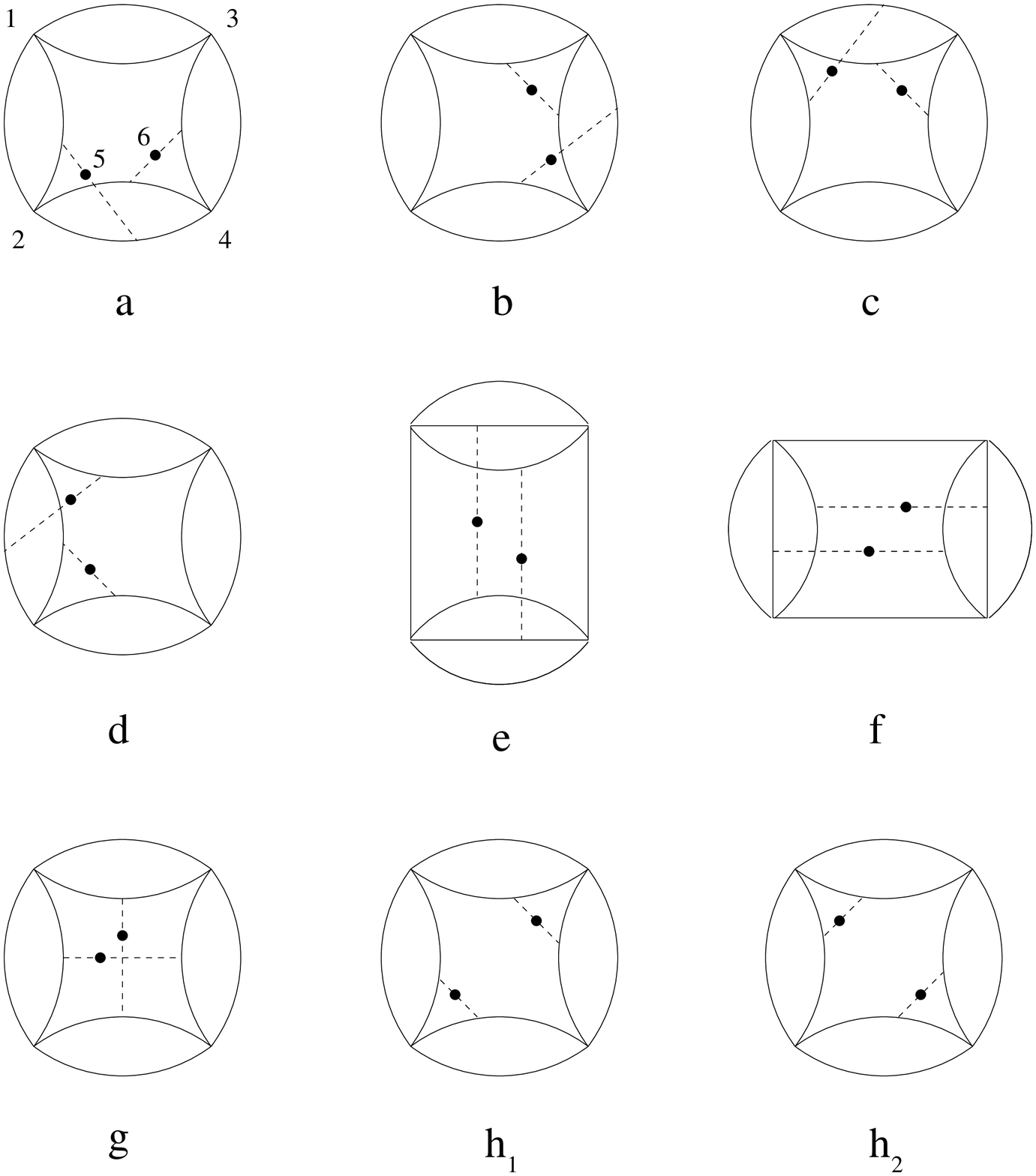}
\end{center}
\begin{center}
Figure 14. Diagrams contributing to the $A_1$ function for $k=4$ case.  
\end{center}
\end{minipage}

\newpage

\renewcommand{\baselinestretch}{0.6}

\end{document}